\documentclass[twocolumn]{aastex62}

\newcommand*\inst[1]{\unskip\hbox{\@textsuperscript{\normalfont$#1$}}}

\newcommand*\institute[1]{
  \begingroup
    \let\and\relax
    \renewcommand*\inst[1]{}%
    \renewcommand*\thanks[1]{}%
    \renewcommand*\email[1]{}%
  \endgroup
  \newcommand{\institutions}{#1}
}%

\let\oldarcsec\arcsec
\renewcommand\arcsec{\oldarcsec\xspace}%

\renewcommand{\ion}[2]{\textup{#1\,\textsc{\lowercase{#2}}}}

\usepackage{latexsym}		
\usepackage{graphicx}		
\usepackage{rotating}		
\usepackage{natbib}  
\usepackage{savesym}
\usepackage{amssymb}
\usepackage{amsmath}
\usepackage{morefloats}
\savesymbol{doublespace}
\usepackage{xspace}
\usepackage{color}
\usepackage{mdframed}
\usepackage{url}
\usepackage{subfigure}
\usepackage{grffile}
\usepackage{import}
\usepackage[utf8]{inputenc}
\usepackage{booktabs}
\usepackage{fancyhdr}

\usepackage[hang,flushmargin]{footmisc}
\usepackage{ifpdf}

\newcommand{\msun}{\ensuremath{M_{\odot}}\xspace}			

\newcommand{\uchii}{\ion{UCH}{ii}\xspace}

\newcommand{\hchii}{\ion{HCH}{ii}\xspace}

\newcommand{\hii}{\ion{H}{ii}\xspace}

\newcommand{\percc}{\ensuremath{\textrm{cm}^{-3}}\xspace}

\newcommand{\persc}{\ensuremath{\textrm{cm}^{-2}}\xspace}

\newcommand{\peryr}{\ensuremath{\textrm{yr}^{-1}}\xspace}
\newcommand{\um}{\ensuremath{\mu \textrm{m}}\xspace}    

\def\ee#1{\ensuremath{\times10^{#1}}}

\newcommand{\perbeam}{\ensuremath{\textrm{beam}^{-1}}\xspace}

\def\eqref#1{Equation \ref{#1}}

\newenvironment{rotatepage}
{}{}

\def\edit#1{{#1}}
\def\change#1{{{#1}}}

\newcommand{\MGPS}{MGPS90\xspace}
\newcommand{\MUSTANG}{MUSTANG-2\xspace}

\newcommand{\nsources}{709\xspace}
\newcommand{\noksources}{279\xspace}
\newcommand{\cmdetections}{119\xspace}
\newcommand{\mmdetections}{240\xspace}
\newcommand{\cmmmnondetections}{34\xspace}
\newcommand{\mmdetectionscmnondetections}{126\xspace}
\newcommand{\cmdetectionsmmnondetections}{5\xspace}
\newcommand{\mmdetectionscmnondetectionscompact}{10\xspace}
\newcommand{\ncompact}{73\xspace}
\newcommand{\nextended}{385\xspace}
\newcommand{\nfilamentary}{251\xspace}

\newcommand{\nhiicand}{5\xspace}\newcommand{\ncompacthiicand}{3\xspace}

\begin{document}
\title{The \MUSTANG Galactic Plane Survey (MGPS90) pilot}
\newcommand{\florida}{\affiliation{\it{Department of Astronomy, University of Florida, PO Box 112055, USA }}}
\newcommand{\nraojansky}{\affiliation{\it{Jansky fellow of the National Radio Astronomy Observatory, 1003 Lopezville Rd, Socorro, NM 87801 USA }}}
\newcommand{\nraonm}{\affiliation{\it{National Radio Astronomy Observatory, 1003 Lopezville Rd, Socorro, NM 87801 USA}}}
\newcommand{\nraocv}{\affiliation{\it{National Radio Astronomy Observatory, Charlottesville, VA 22903 USA}}}
\newcommand{\casa}{\affiliation{\it{CASA, University of Colorado, 389-UCB, Boulder, CO 80309, USA}} }
\newcommand{\eso}{\affiliation{\it{European Southern Observatory (ESO), Karl-Schwarzschild-Strasse 2, Garching 85748, Germany}} }
\newcommand{\gbo}{\affiliation{Green Bank Observatory, 155 Observatory Rd, PO Box 2, Green Bank, WV 24944, USA }}
\newcommand{\upenn}{\affiliation{\it{Department of Physics and Astronomy, University of Pennsylvania, 209 S. 33rd St, Philadelphia PA, 19119, USA }}}
\newcommand{\irya}{\affiliation{Instituto de Radioastronom\'ia y Astrof\'isica (IRyA), UNAM, Apdo. Postal 72-3 (Xangari), Morelia, Michoac\'an 58089, Mexico}}
\newcommand{\asiaa}{\affiliation{Academia Sinica Institute of Astronomy and Astrophysics, P.O. Box 23-141, Taipei 10617, Taiwan}}
\newcommand{\uva}{\affiliation{Department of Astronomy, University of Virginia, 530 McCormick Road, Charlottesville, VA 22904, USA}}
\newcommand{\mcgill}{\affiliation{Department of Physics, McGill University, 3600 University Street Montreal, QC H3A 2T8, Canada}}

\author[0000-0001-6431-9633]{Adam Ginsburg}
\florida

\author{L.~D.~Anderson}
\affiliation{Department of Physics and Astronomy, West Virginia University, Morgantown WV 26506}
\affiliation{Adjunct Astronomer at the Green Bank Observatory, P.O. Box 2, Green Bank WV 24944}
\affiliation{Center for Gravitational Waves and Cosmology, West Virginia University, Chestnut Ridge Research Building, Morgantown, WV 26505}

\author[0000-0002-1940-4289]{Simon Dicker}\upenn
\author[0000-0001-5725-0359]{Charles Romero}\upenn \gbo

\author[0000-0002-8502-6431]{Brian Svoboda} \nraojansky

\author[0000-0002-3169-9761]{Mark Devlin }\upenn
\author[0000-0003-1480-4643]{Roberto Galv\'an-Madrid}\irya
\author{Remy Indebetouw} \nraocv
\author[0000-0003-2300-2626]{Hauyu Baobab Liu} \asiaa
\author[0000-0002-8472-836X]{Brian Mason} \nraocv
\author[0000-0003-3816-5372]{Tony Mroczkowski}\eso

\author[0000-0002-7045-9277]{W.~P.~Armentrout} \gbo
\author{John Bally}
\author{Crystal Brogan} \nraocv
\author[0000-0002-4013-6469]{Natalie Butterfield} \gbo
\author{Todd R.\ Hunter} \nraocv
\author{Erik D.\ Reese} \affiliation{Department of Astronomy, Physics, Engineering, and Computer Science, Moorpark College, 7075 Campus Rd, Moorpark CA 93021 USA}
\author[0000-0002-5204-2259]{Erik Rosolowsky}\affiliation{Department of Physics, University of Alberta, 4-183 CCIS, Edmonton, Alberta, T6G 2E1, Canada}
\author{Craig Sarazin} \uva
\author{Yancy Shirley} \affiliation{Steward Observatory, 933 North Cherry Ave., Tucson, AZ 85721, USA}
\author[0000-0001-6903-5074]{Jonathan Sievers}\mcgill
\author{Sara Stanchfield}\upenn

\begin{abstract}
We report the results of a pilot program for a Green Bank Telescope (GBT)
\MUSTANG Galactic Plane survey at 3 mm (90 GHz), MGPS90.  The survey 
achieves a typical $1\sigma$ depth of $1-2$ mJy\,beam$^{-1}$ with a $9\arcsec$
beam.  We describe the survey parameters, quality assessment process,
cataloging, and comparison with other data sets.  We have identified \nsources\
 sources over seven observed fields selecting some of the most
prominent millimeter-bright regions between $0\deg < \ell < 50\deg$ (total area
$\approx 7.5 \deg^2$).  The majority of these sources have
counterparts at other wavelengths.  By applying flux selection criteria to
these sources, we successfully recovered several known hypercompact HII
(\hchii) regions, but did not confirm any new ones.  We identify
\mmdetectionscmnondetections sources that have mm-wavelength counterparts but
do not have cm-wavelength counterparts and are therefore
candidate \hchii regions; of these, \mmdetectionscmnondetectionscompact are
morphologically compact and are strong candidates for new \hchii regions.
Given the limited number of candidates in the extended area in this survey
compared to the relatively large numbers seen in protoclusters W51 and W49, it
appears that most \hchii regions exist within dense protoclusters.  Comparing
the counts of \hchii to ultracompact HII (\uchii) regions, we infer the \hchii
region lifetime is 16-46\% that of the \uchii region lifetime.  We additionally
separated the 3 mm emission into dust and free-free emission by comparing with
archival 870 \um and 20 cm data.  In the selected pilot fields, most
($\gtrsim80\%$) of the 3 mm emission comes from plasma, either through
free-free or synchrotron emission.
\end{abstract}

\section{Introduction}
Surveys of the Galactic plane in the millimeter regime are essential for measuring
the gas and dust involved in star formation.  Several continuum surveys have covered the
complete plane from the far infrared through 1 mm
\citep{Molinari2010a,Aguirre2011a,Ginsburg2013a,Csengeri2014a,Eden2017a,Elia2017a}.
In the millimeter/submillimeter regime, these surveys have resolution 15\arcsec
or worse.
In the centimeter regime, large-area Galactic plane surveys have been conducted at 4 cm and 
longer wavelengths at resolutions generally $\sim1\arcsec$ or coarser \citep{Giveon2005a,Hoare2012a,Beuther2016a,Medina2019a}.

Emission at 3 mm (90 GHz) consists of a combination of dust, free-free, and synchrotron continuum emission. 
Between 1 mm and 4 cm, there are no existing Galactic plane surveys.  This wavelength
regime represents the global minimum in typical Galactic spectral energy
distributions.  At 3 mm, most dust emission is optically thin; very few regions
have high enough column density $N>3\ee{26}$ \persc on $\sim0.1-1$ pc scales to reach an optical
depth $\tau_{3 \mathrm{mm}}\geq1$.  Similarly, almost all \hii regions exhibit
optically-thin free-free emission  at 3 mm; only the densest of hypercompact
\hii (\hchii) regions are optically thick out to such high frequencies.
Anomalous Microwave Emission (AME) peaks somewhere in the 10-60 GHz
regime and remains a substantial fraction of the total emission on large angular scales out to
$\sim100$~GHz, though so far most observations on smaller ($\lesssim10\arcmin$) scales have
been limited to lower ($<50$~GHz) frequencies  \citep{Dickinson2018a}.

Thermally-emitting dust follows a modified Planck function of typical temperature
\edit{10-30}\,K in Galactic clouds; its intensity therefore peaks near
\edit{1-3}\,THz, \edit{placing the 90 GHz MGPS90 observations firmly on the
Rayleigh-Jeans tail}.  At
90\,GHz, the dust flux density is set by the dust column density $N_d$, the
dust temperature $T_d$,
and the dust opacity $\kappa_\nu$:
\begin{equation}
    S_{\nu, d} \propto \kappa_\nu B_\nu(T_d) N_d,
    \label{eq:dust}
\end{equation}
where $B_\nu(T_d)$ is the Planck function.  The dust opacity as a function of
frequency can be modeled as a power law: $\kappa_\nu \propto \nu^{\beta}$,
where $\beta$ is the dust emissivity index.
Ongoing and future massive star formation is associated with dust emission, and
we expect to see dust emission at 90\,GHz in the MGPS90 fields.  

The flux density from optically thin free-free emission is roughly flat as a
function of frequency, $S_{\nu, ff} \propto \nu^{\alpha}$, where $\alpha=-0.12$
is the spectral index, and we expect almost
all free-free emission to be optically thin at the observed frequency
\citep{Wilson2009a,Condon2007a,Condon2016a}. 

Synchrotron emission generally has a steep negative spectral index and so
decreases in intensity as a function of increasing frequency, $S_{\nu,
synch.}\propto \nu^{\alpha}$, with $\alpha\simeq -1 \mathrm{~to~}
-2$.  At 90\,GHz, we expect to detect synchrotron emission from Galactic
supernova remnants, nonthermal filaments (in the Galactic center), and
extragalactic sources.

The only objects that tend to peak at 3 mm are the most extremely dense and
compact \hii regions.  To reach an optical depth $\tau_{90 \mathrm{GHz}} \sim1$
at 90 GHz, an \hii region must have an emission measure $EM\gtrsim10^{10}$
cm$^{-6}$ pc.  Such high $EM$ is only reached in extremely dense regions
\citep[e.g.][]{Galvan-Madrid2009a}; for example, an $r\sim100$ AU \hii region
would reach $\tau_{90 GHz}\sim1$ at density $n\sim10^7$ \percc
\citep[][]{Wilson2009a,Condon2016a}.  Such compact and dense \hii regions are
expected to be a short phase in the early evolution of massive stars, occurring
shortly after the stars contract onto the main sequence for a brief period
before they expand into less dense, larger \hii regions \citep{Wood1989b}.  A
census of 3 mm peaked, compact sources can provide a measurement of the
actively forming massive star population of the Galaxy, or alternatively by
comparison to other stages, can be used to constrain the lifetime of this early
stage in \hii region evolution.

\MUSTANG \citep{Dicker2014a} is a 215 element bolometer array operating on the
100~m Robert C.\ Byrd Green Bank Telescope\footnote{This material is
based upon work supported by the Green Bank Observatory which is a major
facility funded by the National Science Foundation.} (GBT) with a wide
(75--105~GHz) bandwidth and a 4.25\arcmin\ field-of-view
(fov).\footnote{\url{http://www.gb.nrao.edu/mustang/}}  The TES detectors are 
read out using a microwave multiplexing readout (umux). Typical
observing modes consist of different on-the-fly mapping scans -- either small
daisy scans for arcminute sized targets or larger raster scans in perpendicular
directions used in the data presented in this paper.  Both scan patterns are
designed to maximize cross-linking on many timescales so as to enable the
removal of $1/f$ noise from the instrument and the atmosphere.  In the large
bandwidth of \MUSTANG, line contamination is generally negligible.

We present the first component of an ongoing 3 mm survey with the \MUSTANG
instrument on the GBT with 9\arcsec resolution.   When complete, this survey
will cover most of the northern Galactic plane within $|b|<0.5$.  
This pilot project selected some of the most actively star-forming regions in
the Galaxy to maximize the discovery probability of \hchii regions.  The full
survey will be a blind survey of the Galactic plane.

\section{Observations}

A summary of the reported observations is given in Tables
\ref{tab:observationsummary} and \ref{tab:observations}.

The images from this project are released at \dataset[10.7910/DVN/HPATJB]{https://doi.org/10.7910/DVN/HPATJB}

\begin{table*}[htp]
\centering
\caption{Observation Summary}
\begin{tabular}{llllllll}
    \label{tab:observationsummary}
\edit{Target Object Name}   & \edit{$\ell$ field identifier} & Field Size &     Time  &       Sessions   &  Estimated Noise & $\ell$ offset  & $b$ offset  \\
                     &                         &            &       hr  &                  &  mJy \perbeam    & \arcsec        & \arcsec \\
\hline
\hline
SgrB2      & G01        &$1\deg\times1\deg$ &        1.4 & 02, 03, 04, 05       &        1.7 &        4.0 &        3.3 \\
W33        & G12        &$1\deg\times1\deg$ &        1.0 & 03                   &        1.2 &       -0.4 &        0.3 \\
           & G29        &$1\deg\times1\deg$ &        1.3 & 04,05                &        1.1 &        0.3 &        5.2 \\
W43        & G31        &$1.5\deg\times1\deg$ &        1.5 & 02, 03               &        1.4 &        0.2 &        1.0 \\
G34.26+0.15 & G34        &$1\deg\times1\deg$ &        0.5 & 05                   &        1.2 &        0.5 &        6.9 \\
W49        & G43        &$1\deg\times1\deg$ &        1.0 & 01, 02               &        1.1 &        4.8 &        7.9 \\
W51        & G49        &$1\deg\times1\deg$ &        1.0 & 01                   &        1.2 &        3.5 &        6.3 \\
\hline
\hline
\end{tabular}
\par The $\ell$ and $b$ offsets are the fitted pointing offsets for these
fields compared to 20 cm data; see Section \ref{sec:pointing}.
\edit{The ``Target Object Name'' is the name of the most prominent named object
in the field of view at $\sim90$ GHz, while the ``$\ell$ field identifier'' is
the approximate Galactic longitude center of the field.}
\caption{Observing Session Dates and Lengths}

\begin{tabular}{lllllll}
    \label{tab:observations}
Session Number   & Session Start & Session Length & Beam Peak Major & Minor & Beam Area & $\eta_{peak}$ \\
& & hours & arcsec & arcsec & arcsec$^2$ \\
\hline
\hline
01 & Mar 24 2018 08:00 UT & 3.50 & $9.7\pm0.4$  & $9.1\pm0.2$ & $117\pm9$  & 0.85 \\
02 & Mar 31 2018 07:30 UT & 4.50 & $10.0\pm0.3$ & $9.2\pm0.5$ & $126\pm17$ & 0.83 \\
03 & May 01 2018 06:15 UT & 4.25 & $10.0\pm0.5$ & $9.0\pm0.2$ & $126\pm4$  & 0.81 \\
04 & Jun 15 2018 05:30 UT & 3.25 & $11.1\pm0.7$ & $8.8\pm0.3$ & $127\pm6$  & 0.87 \\
05 & Jan 31 2019 11:45 UT & 2.75 & $10.0\pm0.4$ & $9.3\pm0.3$ & $133\pm11$ & 0.79 \\
\hline
\hline
\end{tabular}

\par
The tabulated times are those in the maps (just in the scans that were used to
make a given map).
\edit{The Beam Peak Major and Minor columns show the average and standard
deviation fit parameters in full-width half-max units of the main peak toward
each of the calibrators.  The Beam Area is the integrated area under the
two-dimensional beam and includes sidelobe contributions.  The $\eta_{peak}$
column measures how much of the beam area is in the central Gaussian beam; it
is the ratio of the area of the Gaussian to the measured beam area.  The
data are peak-calibrated, so this number indicates the fraction ($\sim20\%$)
of the peak flux that is spread into the surrounding larger area ($\sim20\arcsec$).}
In \edit{G34}, only 6 of the constant-latitude scans were completed, so only the bottom
1/3 of map has full cross linking.
\end{table*}

\subsection{Calibration}

A consistent calibration procedure was carried out for each observation.
Known point sources were observed at regular intervals each night.

\begin{enumerate}
    \item A calibration for the detector array, i.e., relative calibration between
        the individual detectors, is found using a skydip and the
        opacity at 90 GHz as given by CLEO (Control Library for Operators and
        Engineers\footnote{\url{http://www.gb.nrao.edu/~rmaddale/CLEOManual/}}) to get
        each timestream into
        antenna temperature.
    \item A map is made in IDL (in azimuth/elevation coordinates) of each
        \change{scan on a calibrator, which is chosen to be an unresolved
        (point-like) source}.  \cite{Romero2020a} describe in detail the
        IDL pipeline for \MUSTANG (MUSTANG IDL Data Analysis System, MIDAS)
        \begin{enumerate}
            \item A single 2-D Gaussian is fit to the point source to
                measure its centroid location.
            \item Fixing the centroid as found above, a double Gaussian is fit.
                The two Gaussian components share a common center; the
                central Gaussian represents the telescope main beam, and the
                second Gaussian represents the first sidelobe of the beam
                response.
            \item \edit{The beam solid angle is calculated both from the fitted
                model parameters and from the sum \edit{of pixel values} within
                a 60\arcsec aperture.  These measurements were consistent, so
                we used the analytically
                derived solid angles from the fitted model parameters.
                These measurements are reported in Table \ref{tab:observations}.
                }
        \end{enumerate}
    \item
        \begin{enumerate}
            \item The peaks of secondary calibrators are normalized by the mean
                flux density for each specific secondary calibrator. These
                peaks are tied to a primary calibrator that is scaled to the
                expected peak in Jy \perbeam.  The expected peak is determined
                from planetary models if a planet is available, or by
                interpolation using available ALMA data
                \citep{vanKempen2014a,Fomalont2014a} if no planet with a
                suitable flux model is accessible.\footnote{We use
                standard ALMA calibrators from the GridCal program.  See
                \url{http://www.alma.cl/~ahales/cal_survey/plots/calsurvey_monitoring_B3.html} and 
                \url{https://almascience.eso.org/sc/}.}
                The scaling is
                linearly interpolated between calibration scans.
            \item Conversion to Rayleigh-Jeans brightness temperature (in K;
                see e.g. \citealt{Condon2016a}) accounts for the beam solid
                angle. As such, the beam solid angles are interpolated between
                scans.
        \end{enumerate}
    \item Calibration to Jy, conversion from Jy \perbeam to Kelvin, opacities,
        and pointing offsets are recorded in an IDL save file and are
        applied to the processing of the \change{time ordered data taken on the
        science target (in this case, scans of the Galactic plane)}.
\end{enumerate}

The absolute accuracy of these calibrations is about 10\%.  Some of this
uncertainty is from the extrapolation in time and frequency of the ALMA sources
(the ALMA band is different from \MUSTANG but there are measurements at
$\sim~100$ and 91 GHz), some is the error in the point source fluxes from ALMA,
and some is from our knowledge \change {of the optical depth $\tau_{90 GHz}$}
during the observations (for which we use archival weather data and models of
the atmosphere).

\subsection{Map Making}
Maps \edit{of the science fields} were made using \MUSTANG's MINKASI (Sievers
et al. in prep) data reduction pipeline which is based on the maximum
likelihood pipeline written for the \change{Atacama Cosmology Telescope
\citep[ACT;][]{Dunner2013a}}.
We used smoothed power spectra from a singular value decomposition (SVD) of the
data on a scan by scan basis to obtain a noise model.
This model does not work well if there are strong sources.
By subdividing timestreams and taking power spectra of each segment, it is
possible to identify power spectra taken from parts of the timestreams with
strong sources as there is a significant increase in the signal band
($\sim$0.1--15~Hz). These regions are flagged and an average power spectrum
\edit{is} calculated from the median of the remaining segments.

We followed an iterative process to obtain the best maps.  A map is made, the
result then clipped at some level above any artifacts in that iteration and the
results subtracted from the timestreams.  In each loop, the clipping level was
reduced and the noise model recalculated.  In the last loops (in which all
strong signal should have been removed) the full SVD noise model could be used
(which tended to give better results on faint features). For W33, three
iterations produced optimal results; the other regions required more
iterations.

For some fields, notably G34, we only obtained scans in one direction.  Future
observations filling in the orthogonal scan direction will be needed to eliminate
the resulting scan-direction striping features.

The map making process assumes the mean incoming intensity is zero.  This
assumption encodes a large angular scale filter such that angular scales larger
than $\sim4.25$\arcmin are not present in the data.  This filtering is visible
as negative bowls in the images, especially in the Sgr B2 / Galactic Center
field.

\begin{figure*}[htp]
\includegraphics[width=17cm]{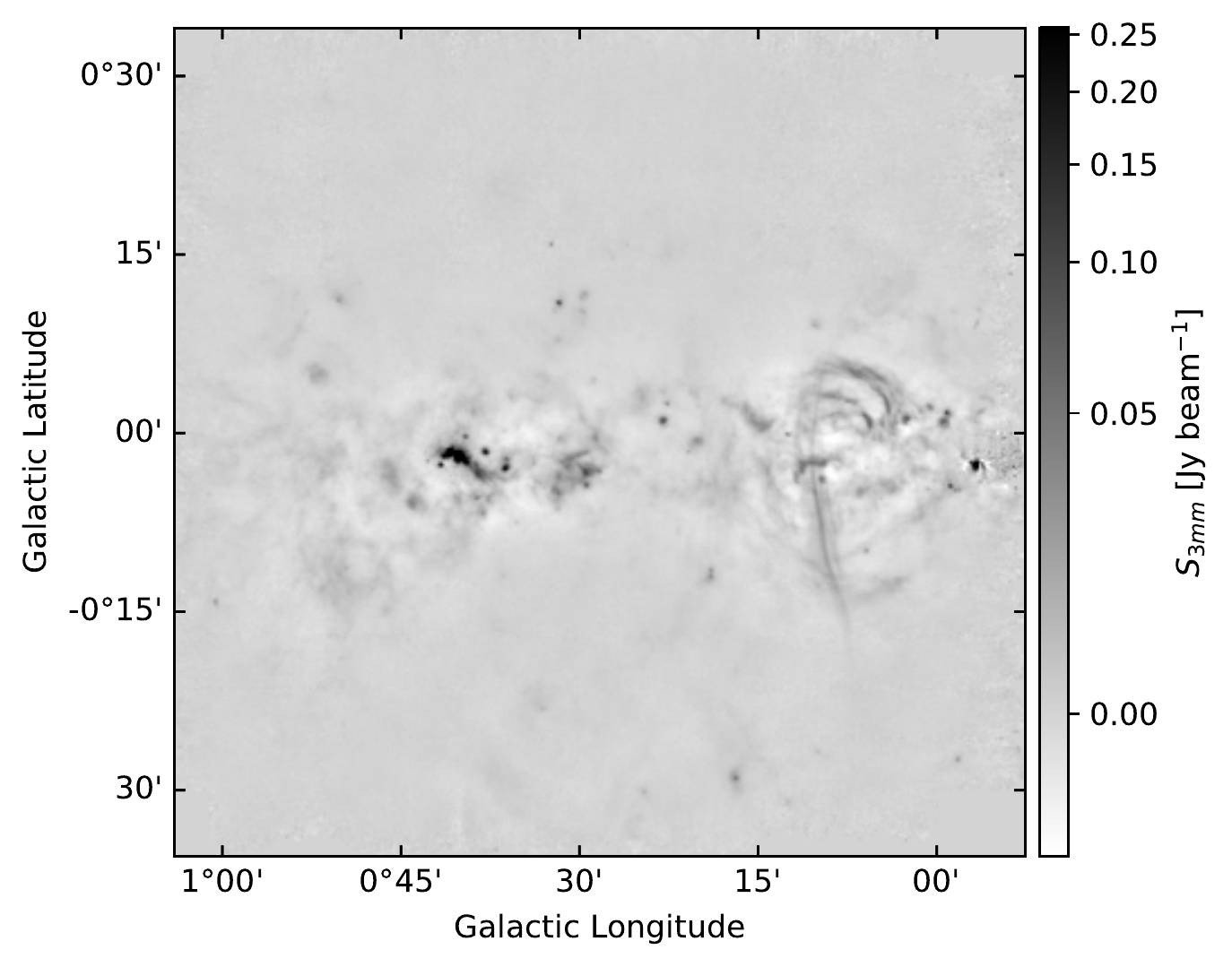}
\caption{\MUSTANG image of the G01 field, centered on Sgr B2.
}
\label{fig:g01overview}
\end{figure*}

\begin{figure*}[htp]
\includegraphics[width=17cm]{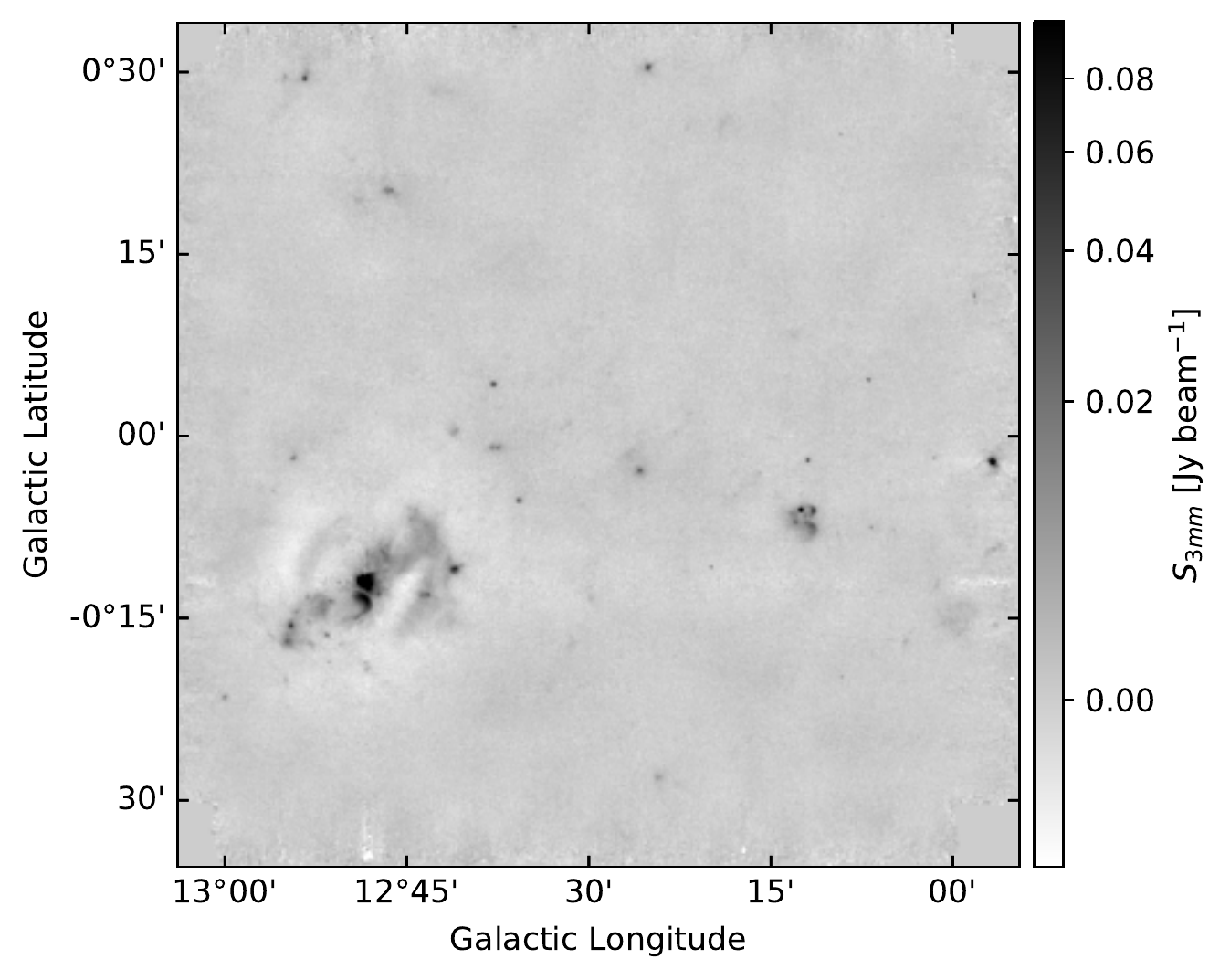}
\caption{\MUSTANG image of the G12 field, including the W33 star-forming region.}
\label{fig:g12overview}
\end{figure*}

\begin{figure*}[htp]
\includegraphics[width=17cm]{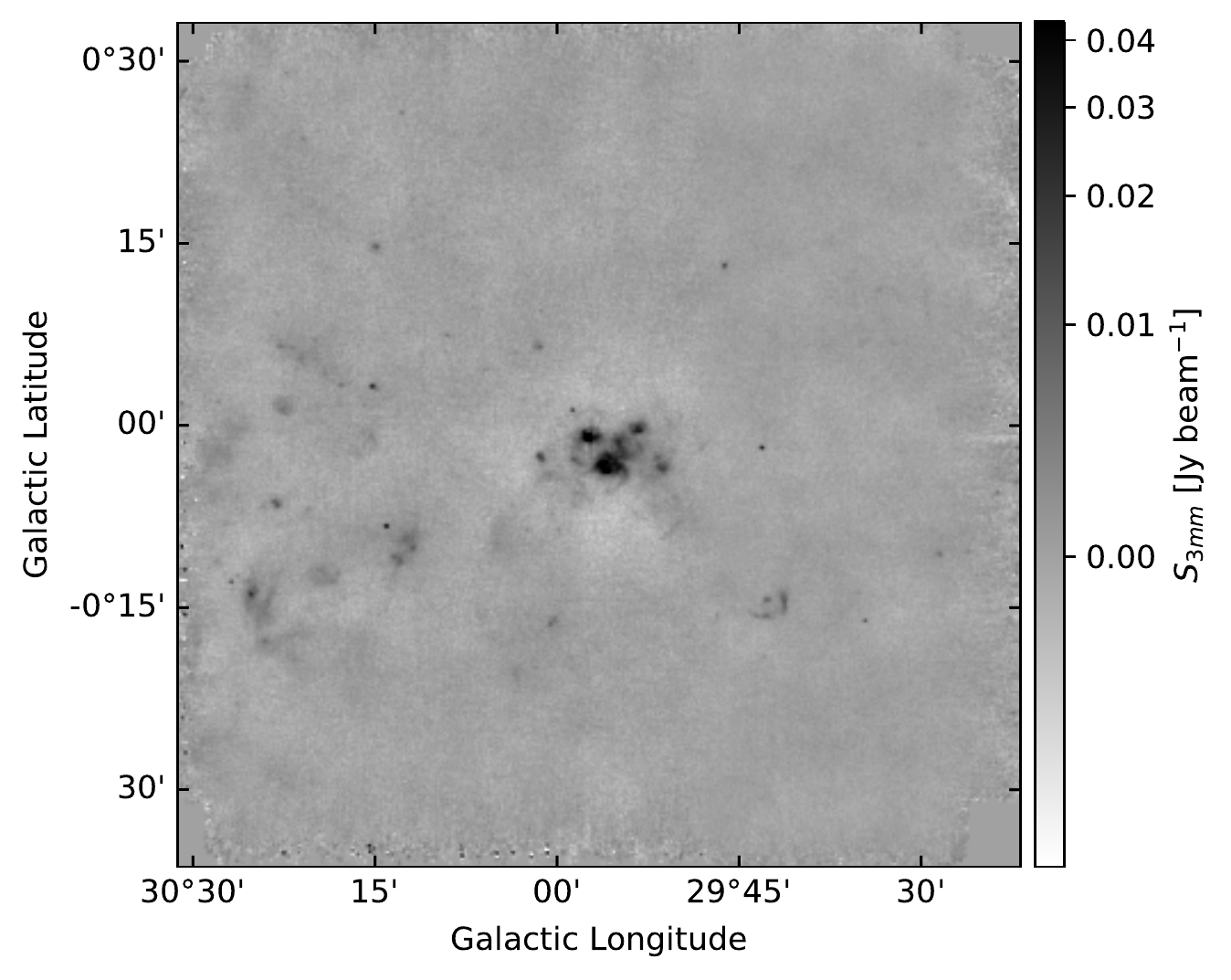}
\caption{\MUSTANG image of the G29 field.}
\label{fig:g29overview}
\end{figure*}

\begin{figure*}[htp]
    \includegraphics[width=17cm]{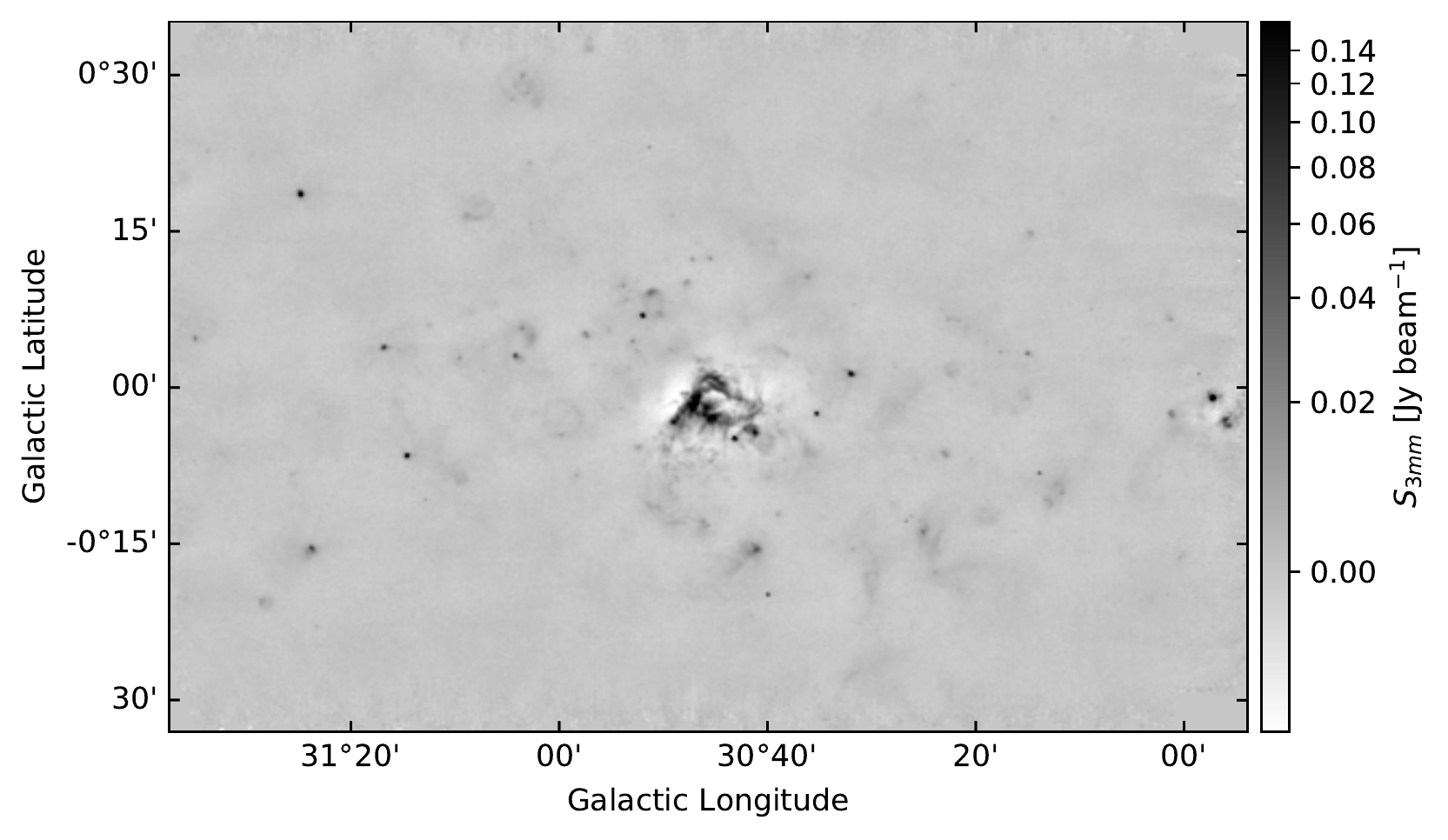}
\caption{\MUSTANG image of the G31 field containing W43.
}
\label{fig:w43overview}
\end{figure*}

\begin{figure*}[htp]
\includegraphics[width=17cm]{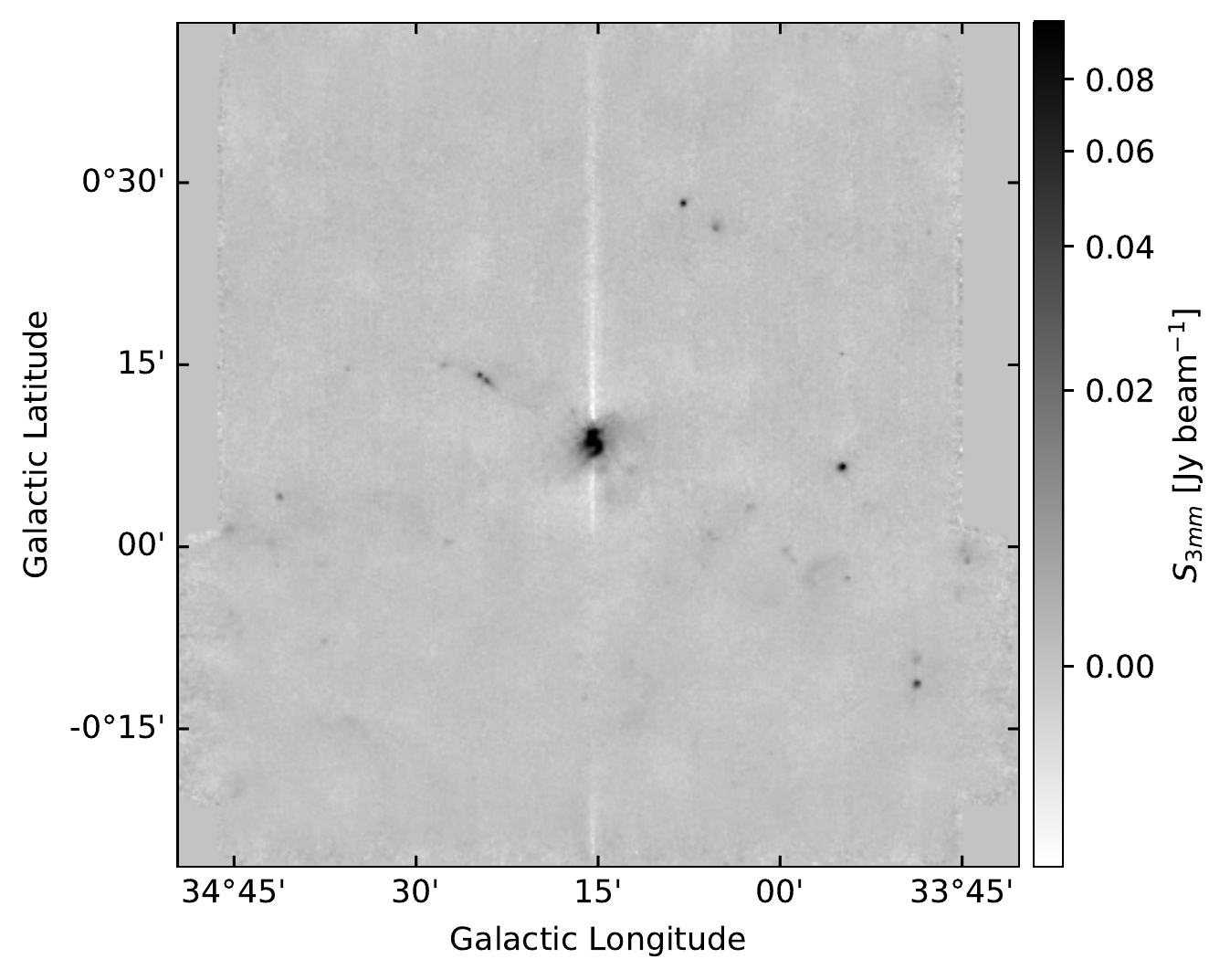}
\caption{\MUSTANG image of the G34 field.  Above $b\gtrsim 0\deg$,
horizontal cross-scans have not been obtained; the vertical streak seen
at $\ell=34.25\deg$ is a consequence of these missing data.}
\label{fig:g34overview}
\end{figure*}

\begin{figure*}[htp]
\includegraphics[width=17cm]{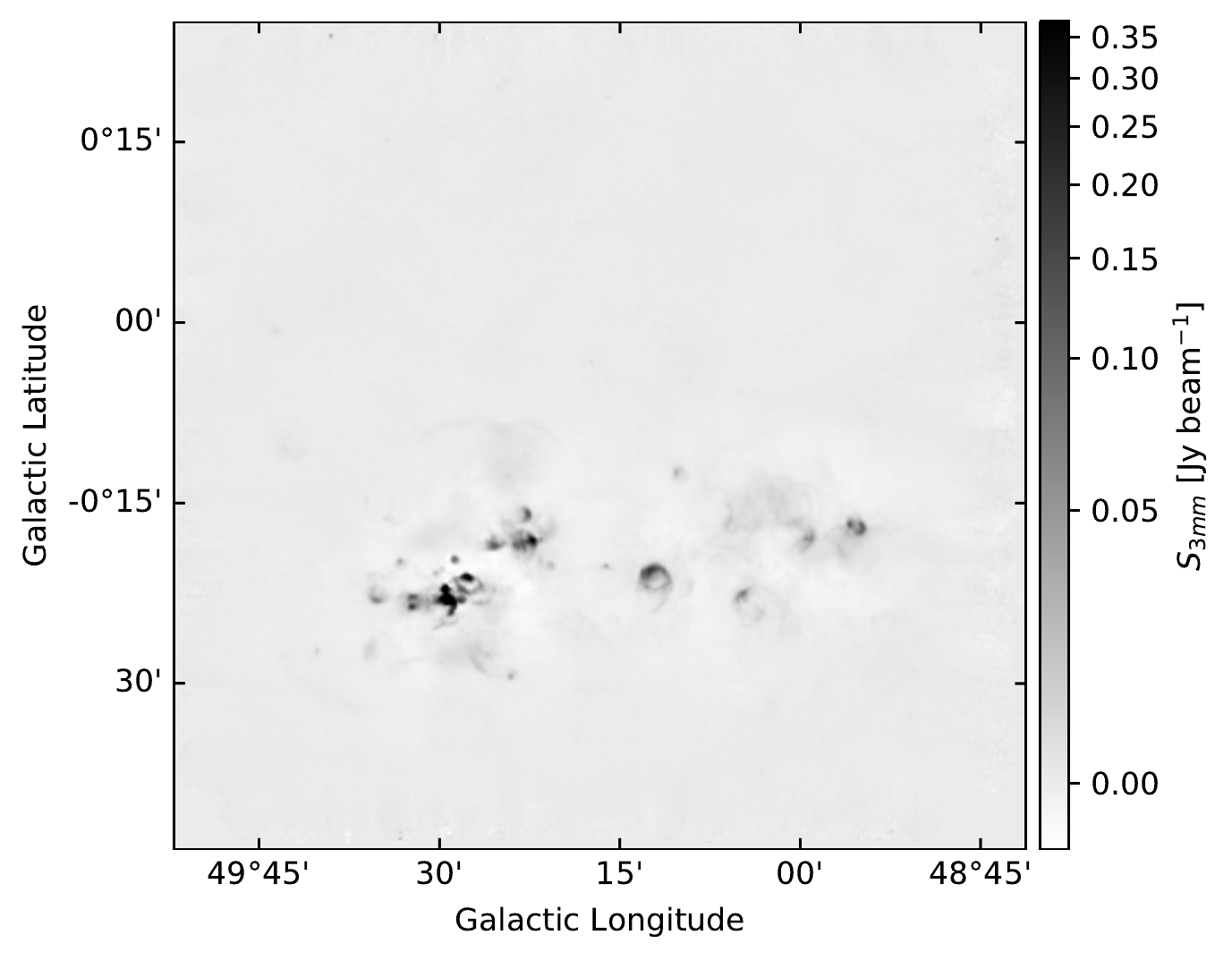}
\caption{\MUSTANG image of the G49 field, which contains the W51 star-forming
complex.}
\label{fig:g49overview}
\end{figure*}

The processed images are shown in figures
\ref{fig:g01overview}-\ref{fig:g49overview}.

\subsection{Sensitivity and beam size}

The effective beam size in the delivered maps is the convolution of the
intrinsic FWHM = 8.1\arcsec beam with a FWHM = 4\arcsec Gaussian kernel,
resulting in a 9\arcsec beam.  \edit{This smoothing suppresses
sub-beam-scale noise at a modest cost in beam area.}
The errors per beam reported in Table
\ref{tab:observationsummary} correspond to these smoothed images.

\subsection{Pointing Accuracy}
\label{sec:pointing}
Several corrections to the raw timestream data were required to produce maps.
Individual scans were noted to have point sources shifted by up to half a beam
($\sim4\arcsec$), indicating a timing error between the \MUSTANG pointing data
and the true telescope pointing.  To ensure that point sources were coincident
in the maps, scans were cross-correlated with a first-iteration map, then
assigned a new timing offset.  The timing errors ranged from $\sim10$ to $30$
milliseconds, \edit{corresponding to angular scales of $\approx1-3\arcsec$ at our
scan rate of $\approx90$ arcseconds/second (scan speeds vary during an observation)}.
Additional half-beam \edit{timing-related} pointing errors were noted in some
individual scans, resulting in additional streaking artifacts in the data.
Most of these issues disappeared after smoothing the data with the 4\arcsec
kernel.

We compared the \MUSTANG maps with 20 cm images from the MAGPIS Galactic Plane
survey \citep{Helfand2006a} and from other sources
\citep{Mehringer1994a,Yusef-Zadeh2004a} to measure pointing offsets, since
these images showed the closest morphological match to the MGPS90 data.
However, there are substantial regions in each field, particularly the Galactic
center, that are synchrotron-dominated at 20 cm and have no corresponding
features at 3 mm; we masked out these features.  We use the
\texttt{image-registration}\footnote{\url{http://image-registration.rtfd.org}}
toolkit to cross-correlate the \MUSTANG images with the 20 cm
images and use a Fourier-domain upsampling approach to obtain
sub-pixel positional offsets.  We were not able to measure statistical uncertainties
on these offsets, but correcting the images for the offsets resulted in smaller
visual residuals in the difference images shown in Section \ref{sec:freefree}.
The measured offsets are reported in Table \ref{tab:observations} and show the
offset of the 20 cm data with respect to
the \MUSTANG data.  \edit{The mean and standard deviation offset from the 20 cm
data are $\Delta\ell=1.8\pm2\arcsec$ and $\Delta b=4.4\pm2.7\arcsec$,
respectively.}

In several cases, the measured offset is comparable to the \MUSTANG beam.  We
therefore correct these images for the offset, assuming the VLA 20 cm data
have correct pointing.  The original pointing centers are recorded in the FITS
headers of the published images with names \texttt{CRVALnA} so that the original
pointing centers can be used if needed.

\subsection{Effective Central Frequency}\label{sec:eff_freq}
The \MUSTANG bandpass filter is approximately flat over the range 75 to 105
GHz, though including surface inaccuracies via the Ruze formula, the effective
sensitivity declines by about a factor of three over this range.  We multiplied
the bandpass filter by power law flux density distributions with
$S_\nu\propto\nu^{\alpha}$ to obtain the true effective central frequency of
the bandpass for these assumed continuous distributions.  They are reported in
Table \ref{tab:centralfreq}.

\begin{table}[htp]
\centering
    \caption{Central Frequencies}
\begin{tabular}{lcc}
    \label{tab:centralfreq}
$\alpha$ & Frequency & Wavelength\\
         & (GHz)       & (mm) \\
\hline
0.0 & 87.85 GHz & 3.413 mm\\
0.5 & 88.23 GHz & 3.398 mm\\
1.0 & 88.62 GHz & 3.383 mm\\
1.5 & 89.02 GHz & 3.368 mm\\
2.0 & 89.41 GHz & 3.353 mm\\
2.5 & 89.80 GHz & 3.338 mm\\
3.0 & 90.19 GHz & 3.324 mm\\
3.5 & 90.58 GHz & 3.310 mm\\
4.0 & 90.96 GHz & 3.296 mm\\
\hline
\end{tabular}
\par The central frequencies are computed by integrating the first moment of a
power-law source function $S(\nu) = \nu^{\alpha}$ over the \MUSTANG
bandpass including the effect of surface errors using the Ruze formula
with an RMS surface accuracy 230 \um \citep{Frayer2018a}.

\end{table}

\subsection{Combination with Planck data\label{sec:feather}}
The largest angular scale recovered by the \MUSTANG data pipeline is
approximately 4.25\arcmin.  Large angular scale structure is therefore missing.
To recover those missing scales, we combine the \MUSTANG data with
Planck 100\,GHz data (with an effective central frequency of 104.225\,GHz
assuming a spectral index $\alpha=3$) scaled to \change{an adopted} central frequency
of 90.19\,GHz for \MUSTANG, \change{as appropriate for $\alpha=3$ (see Section
\ref{sec:eff_freq})}.  We use a simple \texttt{feather} procedure
\citep{Cotton2017b} as implemented in the \texttt{uvcombine}
\footnote{\url{https://github.com/radio-astro-tools/uvcombine}} \edit{python}
package.
Planck's spatial resolution is $\approx10\arcmin$, substantially larger
than the largest scale recovered in the MGPS data, so intermediate-scale
structures (4-10\arcmin) are likely recovered poorly.
These data are not used in the analysis in this paper, but the
FITS images are provided in the data repository.

\section{Compact Source Catalogs}

We use
\texttt{astrodendro}\footnote{\edit{\url{https://dendrograms.readthedocs.io/en/stable/}}}
via the
\texttt{dendrocat}\footnote{\edit{\url{https://dendrocat.readthedocs.io/en/latest/}}}
wrapper to extract a
catalog of compact structures.  In brief, \texttt{astrodendro} catalogs
hierarchically nested signal, effectively cataloging contoured regions.  For
the catalog described here, we included only the most compact structures, which
are the `leaves' in the catalog hierarchy.

To select primarily robust compact sources, we filter the images to reject
scales $>45\arcsec$ prior to cataloging.  We use a $4 \sigma$ flux threshold
and minimum of 100 pixels as the dendrogram parameters; the pixel scale is
1$\arcsec$/pixel, so our minimum object size is $\sim1/2$ of beam area.  We
then reject sources with a peak signal-to-noise ratio less than 5, where we
used the average noise level across the field.  We report the noise level
estimated using the median absolute deviation scaled to the standard deviation
for each field in table \ref{tab:observationsummary}.

The resulting catalog includes all of the significant pointlike sources in each
field of view.  However, this catalog also includes components of extended
emission that had peaks that met the threshold criteria but are not distinct sources.
The extended objects are a particularly prominent component of the Galactic center
field.

To eliminate some of the extended structures, we then fit Gaussian profiles to
each of the dendrogram-identified sources using the \texttt{gaussfit\_catalog}
package\footnote{\url{https://github.com/radio-astro-tools/gaussfit_catalog/}}.
Profiles were fitted to the original, unfiltered data.  Profiles were
restricted to have major and minor axes FWHM$<27\arcsec$, restricting the fits
to be within a factor of three of the beam size.  \edit{Sources substantially larger
than this likely have measured integrated intensities attenuated by the filter
function of the data acquisition and reduction pipeline; however, the full
spatial transfer function of \MUSTANG has not yet been measured.} Fits were
performed to a
30\arcsec radius around each source.  If a second source was present in that
radius, it was masked out with a single-beam-FWHM circle.  

A total of \nsources sources were identified across the seven fields.
\edit{Of these, the majority, \nextended were extended and round ($\sigma_{maj} >
14\arcsec$), and an additional \nfilamentary had both long aspect ratios
$\sigma_{maj}/\sigma_{min} > 1.5$ and were extended ($\sigma_{maj} >
14\arcsec$)}.
Only \ncompact sources were compact \edit{($\sigma_{maj} < 14\arcsec$)}.  Note that any confused or
clustered sources, e.g., two compact sources within $\sim5-20\arcsec$ of one
another, would likely be classified as extended.

The full catalog is available on the project source code
repository.\footnote{The January 8, 2020 version is at
\url{https://github.com/keflavich/MGPS/blob/c81af46342d057b75c372d298074084415dcdf08/tables/concatenated_catalog.ipac}.}
A complete description of the catalog columns and an excerpt from the catalog
are both shown in Appendix \ref{appendix:Catalog}.

\subsection{Catalog cross-matching}
\label{sec:catalogmatching}
We cross-match the resulting catalog with the catalogs listed in Table \ref{tab:otherdata}.  
Matches in these catalogs are included if there is a source within 10\arcsec
(approximately the \MUSTANG beam FWHM) of the MGPS catalog entry.

\begin{table*}[htp]
\centering
\caption{Comparison Data Set Summary}
\begin{tabular}{lllll}
\label{tab:otherdata}
Name                 & Wavelength(s)  &  Angular Resolution  & Approximate Sensitivity$^*$  & References \\
                     & \um            &  \arcsec             & mJy \perbeam  &  \\
\hline
Spitzer GLIMPSE      & 3.6--8.0       &  2                   & -  & \citet{Churchwell2009a} \\
Spitzer MIPSGAL      & 24             &  6                   & -  & \citet{Gutermuth2015a} \\
Herschel Hi-GAL       & 70--500        &  6--36               & 20-85  & \citet{Molinari2016a,Elia2017a} \\
APEX-Laboca ATLASGAL & 870            &  20                  & 70  & \citet{Urquhart2014b} \\
CSO-Bolocam BGPS     & 1100           &  33                  & 50  & \citet{Rosolowsky2010a}\\
                                                             & &&& \citet{Ginsburg2013a} \\
GBT-\MUSTANG MGPS90  & 3274           &  9                   & 1-2  & This work \\
\hline
                     & cm             &                      &   & \\
\hline
MAGPIS               & 6              & 4                    & 2.5  & \citet{Giveon2005a} \\
                                                             & &&& \citet{Helfand2006a} \\
CORNISH              & 6              & 1.5                  & 2.5  & \citet{Hoare2012a} \\
MAGPIS               & 20             & 5                    & 2  & \citet{Giveon2005b} \\
\hline
\end{tabular}
$^*$ When no matching entry was found, we adopted the listed value as a
$1-\sigma$ upper limit on the source flux when plotting SEDs.  However, all of these
surveys have significantly varying point source sensitivity at different locations,
so these limits should be treated as very loose.  The GLIMPSE and MIPSGAL
\edit{catalog} data were not cross-matched, so no upper limit was used \edit{in
SED fitting}, but we used \edit{images} from these surveys to produce cutout
images for morphological comparison \edit{(see Figures \ref{fig:g29pn} to
\ref{fig:g34hchii})}.  The Hi-GAL sensitivity and angular resolution values
both rise from 70 to 500 \um.  \edit{Note that the PACS 70\um beam in the Hi-Gal
data set is asymmetric, $\sim6\times12\arcsec$ \citep{Molinari2016a}.}

\end{table*}

Of the \nsources total \MGPS sources, \change{\noksources passed our selection criteria that
the peak signal-to-noise ratio in the source was $SNR>5$ and the peak signal was at least
twice that of the background, $S_{peak} > 2 S_{background}$.}.
Of those, \mmdetections had millimeter/submillimeter
matches (Herschel 70-500 \um, LABOCA 870 \um, or Bolocam 1.1 mm), \cmdetections 
had centimeter-wavelength matches (6 cm or 20 cm), and \cmmmnondetections
had no match in the millimeter or centimeter catalogs.
There were \mmdetectionscmnondetections sources cross-matched at shorter
wavelengths but not at longer wavelengths, \edit{and \cmdetectionsmmnondetections
with long-wavelength matches but no short-wavelength}.

\begin{figure*}[htp]
    \includegraphics[width=17cm]{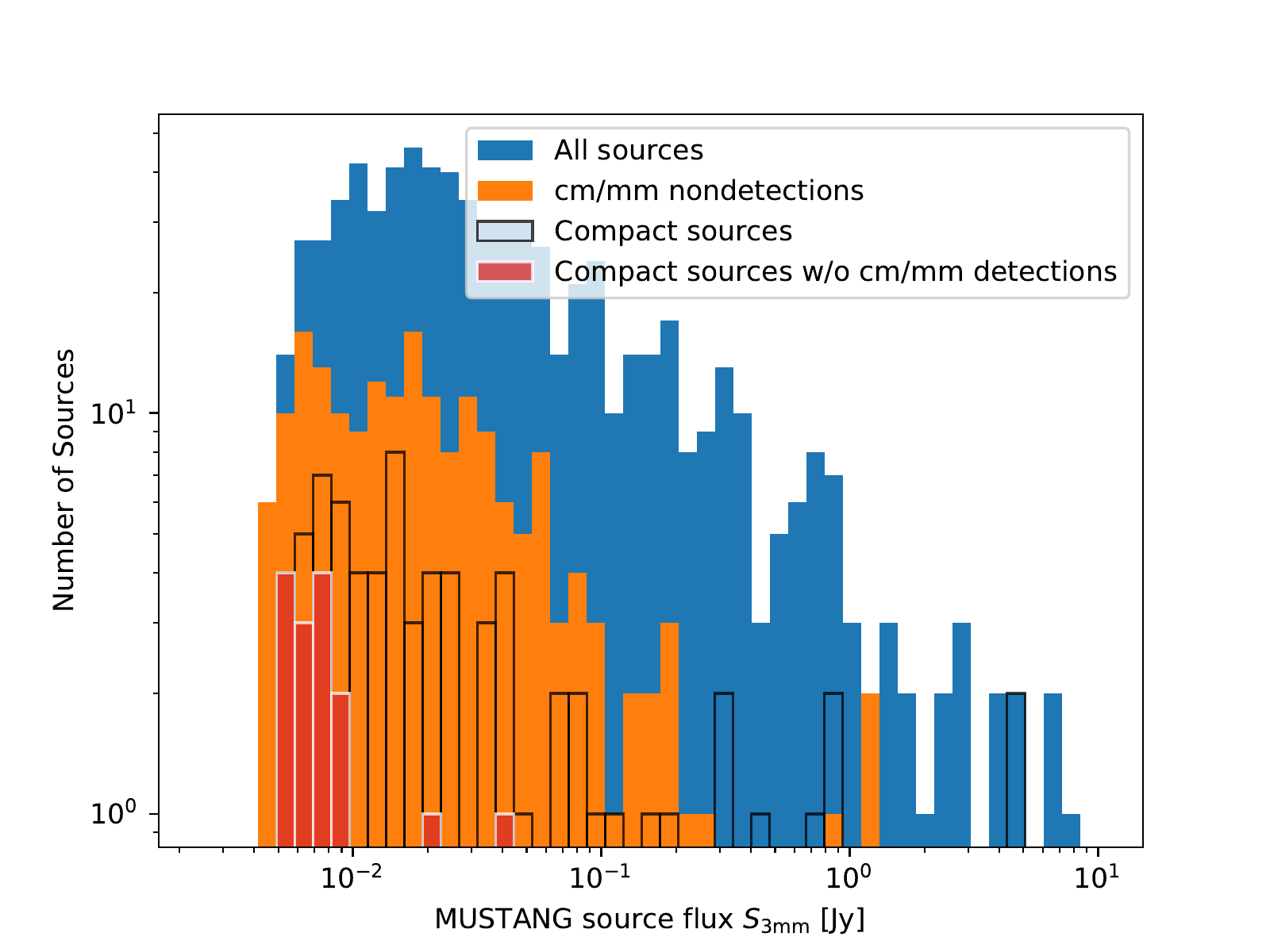}
    \caption{Histogram of the source catalog.  Blue shows all sources,
    while orange (overlaid \edit{as a foreground layer}) shows only those sources that have
    no matches at cm or mm wavelengths in the searched surveys.}
\label{fig:fullcathist}
\end{figure*}

Figure \ref{fig:fullcathist} shows the histogram of \MUSTANG-measured fluxes
in the catalog.  Because the typical noise level was $\sim1$ mJy, the catalog
has few sources below 5 mJy.  The overlaid histogram shows the subset of
the sample with no detections at other wavelengths; this subset is much
fainter than the overall distribution, suggesting that the majority
of these sources were either below the detection limit or the confusion
limit of the other surveys.

\subsection{\hchii region identification}
\label{sec:hiireg}
One of the aims of this survey is to identify the youngest high-mass
protostars.  Candidates are those sources with little to no mid-infrared emission
and very compact, optically thick (hypercompact) \hii regions.

Massive stars form in the middle of ultra-dense cores undergoing gravitational
collapse, leading to an accretion rate of order $\dot{M} \sim 10^{-3}$ \msun
\peryr such that a 100 \msun star takes about $10^5$ years to accrete its mass.
As the star contracts onto the main sequence it starts to ionize its
environment to create an \hchii region.  For a sufficiently dense accretion
flow, the Str{\"o}mgren radius of the \hchii region is bound by the gravity of
the star, with a radius $R_G \sim$50-100 AU
\citep{Keto2002a,Keto2003a,Keto2007a}.  Such gravitationally bound \hchii
regions are optically thick at centimeter wavelengths and therefore emit as
blackbodies at wavelengths $\lambda\gtrsim$3~mm, with 
\begin{equation}
    S_\nu=21~\textrm{mJy} \left(\frac{d}{5~\textrm{kpc}}\right)^{-2} \left(\frac{R}{100~\textrm{AU}}\right)^2 \left(\frac{\nu}{90~\textrm{GHz}}\right)^2
\end{equation}
which is only
0.06 mJy at $\nu=5$ GHz, and therefore below the detection limit of many
existing surveys; they are certainly unremarkable sources at long wavelengths.
\hchii regions can be distinguished from older
ultracompact (\uchii) regions by their  bright 90\,GHz emission and faint emission at 5\,GHz and lower
frequencies.  Sources
with free-free emission that peaks at or just below 3 mm represent the youngest
high-mass YSOs.  The dense cores surrounding these sources will be bright
in the millimeter regime, since they will have high dust column densities and
temperatures.

We therefore select candidate HC\hii regions
as those fitting either of these criteria:
\begin{enumerate}
    \item $S_{3 \mathrm{mm}} > 1.75~S_{6\mathrm{cm}}$.  
        This requirement selects free-free sources that have $\tau_\textit{ff}=1$
        at ${\lambda={6~\mathrm{cm}}}$.  It corresponds to an emission
        measure ${EM=7\ee{7}~\mathrm{cm}^{-6}~\mathrm{pc}}$.
    \item
        The source is not detected at 6 and 20 cm, is detected at 1.1~mm, and has
        \begin{equation}
            \frac{S_{3\mathrm{mm}}}{S_{1.1\mathrm{mm}}} > \left(\frac{3.28~\mathrm{mm}}
            {1.11~\mathrm{mm}}\right)^{-\alpha} = 0.039
        \end{equation}\\
        where $\alpha=3$ is the spectral
        index for optically thin dust with an opacity index $\beta=1$.  This
        requirement selects dust-detected sources in which there is some
        indication of an excess of free-free emission over pure dust emission
        at 3~mm.
        \hchii regions that are optically
        thick up to $\sim3$ mm, those that are extremely compact and dense,
        are below the detection threshold of the centimeter surveys
        \citep[$\approx2.5$ mJy at 6 cm;][]{Giveon2005a,Hoare2012a}.
\end{enumerate}

These criteria provide a small sample of \nhiicand candidate \hchii regions
across the seven target regions.  Only \ncompacthiicand of these candidates
were morphologically compact. This sample consists of known
ultracompact or \hchii region clusters \citep[three are parts of
W49A, which contains 12 sources that can be classified as \hchii
regions;][]{De-Pree1997a},
the \hchii region G34.257+0.153, and the OH/IR star G30.944+0.035 \citep{Wilson1972a}.
The ten known \hchii regions in W51 \citep{Ginsburg2016b} were not recovered because
they are blended, in the 9\arcsec \MUSTANG beams, with more diffuse \hii regions.

However, the majority of sources in our catalog do not have
centimeter-wavelength detections and therefore were not
eligible to be selected based on criterion 1 above.  The BGPS 1.1 mm data,
which have only 30\arcsec resolution, could be affected by confusion (source blending)
and therefore be too bright for a 3 mm excess to be detected, preventing
selection by criterion 2.

While we would expect some \edit{free-free} excess \edit{at 90 GHz} above the
dust emission \edit{extrapolated from 1.1 mm in} dusty \hchii regions, it is
plausible that the excess is not enough to modify the spectral index to meet
our selection criterion 2.  Sources that have
millimeter detections (since they must be surrounded by gas and dust) and not
centimeter detections therefore remain candidate \hchii regions.  This large
sample of \mmdetectionscmnondetections additional candidates, especially the
\mmdetectionscmnondetectionscompact that are compact, are interesting
candidates for future deep centimeter observations.

Several well-known \hchii regions were excluded from these selection
criteria.  The \hchii regions in W51, including the W51e cluster and W51d2
\citep{Ginsburg2016a}, those in W49 \citep{De-Pree1997a}, and those in Sgr B2
\citep{De-Pree1998a} are confused, residing in the same beams as other
high-mass stars at different evolutionary states.  G34.257+0.153 includes a pair
of \hchii regions but less other surrounding emission, so it did pass our selection
criteria \citep{Sewilo2004a,Avalos2006a}.  MGPS90 is clearly capable of detecting
\hchii regions that are not in dense protoclusters.

\subsection{Constraints on \hchii lifetimes}
To estimate the relative lifetime of the hypercompact and ultracompact phases,
we compare the number of \hchii candidates to the number of detected \uchii
regions from the CORNISH survey \citep{Kalcheva2018a}.  \citet{Wood1989a}
seeded the idea that \uchii lifetimes may be substantially longer than expected
for a freely expanding Str\"omgren sphere $t_{\textrm{fe}} 
\sim 4\ee{4}$ years,
but the improved sample of \citet{Kalcheva2018a} suggests that the discrepancy
is not so large.  In any case, we adopt a loosely estimated \uchii lifetime
\change{within the range} $4\ee{4}~\mathrm{yr} < t_{\uchii} <
4\ee{5}~\mathrm{yr}$.

In the observed regions, the CORNISH survey detected 73 \uchii regions.  Over
the same area, our sample includes \mmdetectionscmnondetectionscompact compact
\MUSTANG sources with no centimeter detections, which are our additional
candidates from \S \ref{sec:hiireg}, and four previously-known \hchii regions.
W51 contains 10 and W49A contains up to 12 additional \hchii region candidates
when viewed at high resolution \citep{De-Pree1997a,Ginsburg2016a}.  The
inferred lifetime of \hchii regions, using a sample size of 12-34 \hchii's in
the \MGPS fields, is therefore $16-46\%$ that of \uchii regions, or $6\ee{3}
\mathrm{yr} < t_{\hchii} < 2\ee{5} \mathrm{yr}$.  A more complete assessment
from the larger survey may more tightly constrain these values.

Furthermore, though, the relatively small number of new candidates (only
\mmdetectionscmnondetectionscompact) compared to the large numbers in compact
regions suggests that \hchii regions
form primarily in, or live longest in, clustered regions.  This high production
of \hchii regions in dense protoclusters can be either because more high-mass stars
form there, indicating an overall higher population, or because the gas density is
higher, allowing the \hii regions to remain in the hypercompact phase for a longer
period before expanding into \uchii or diffuse \hii regions.

\begin{figure*}[htp]
\includegraphics[width=17cm]{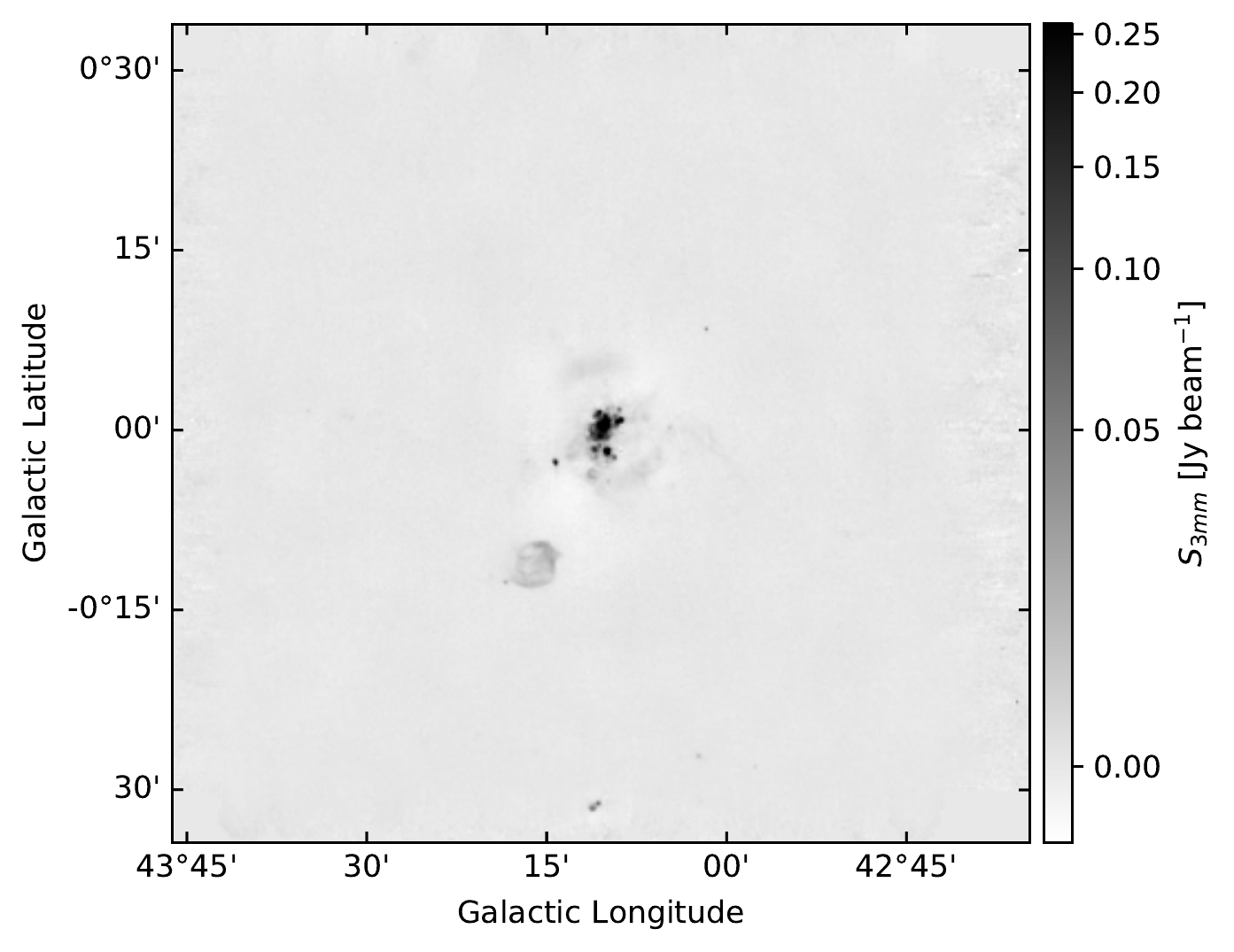}
\caption{\MUSTANG image of the G43 field, which contains the 
W49A star-forming complex (center) and the W49B supernova remnant (just southeast
of center).}
\label{fig:g43overview}
\end{figure*}

\subsection{Representative SEDs of selected sources}
\begin{figure*}[htp]
\includegraphics[width=17cm]{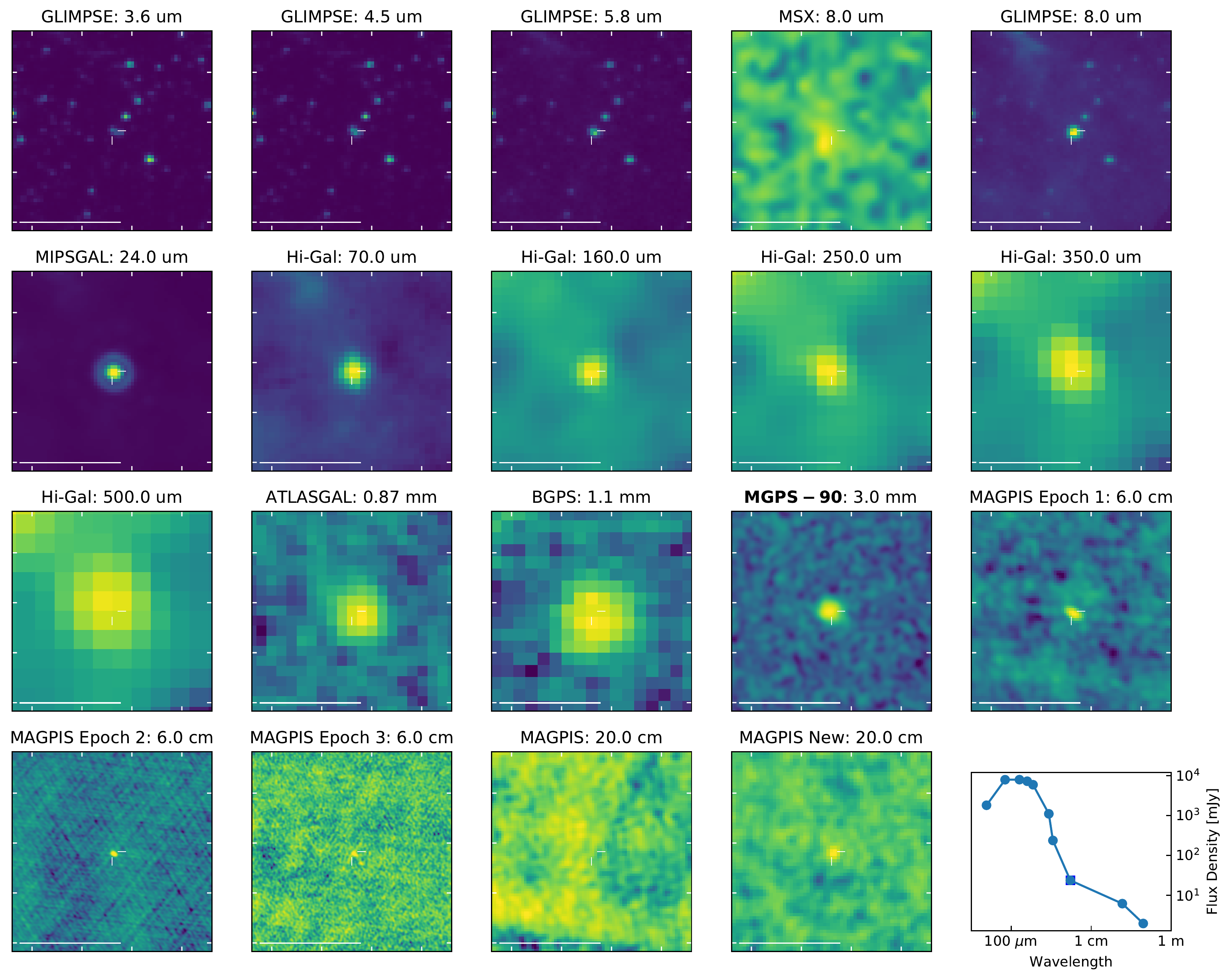}
\caption{SED of source G29.578-0.268, a candidate planetary nebula.  The
individual panels show cutouts from various surveys hosted on the MAGPIS cutout
service (\url{https://third.ucllnl.org/gps/}) and the Herschel Hi-GAL cutout
service (\url{https://tools.ssdc.asi.it/HiGAL.jsp}); individual survey
references are in Section \ref{sec:catalogmatching}.  The white bar in the
lower-left is a 1\arcmin\ scalebar.  Short wavelengths ($<24$ \um) are not
included in the SED, the rest are included in the cross-matched catalog.
}
\label{fig:g29pn}
\end{figure*}

\begin{figure*}[htp]
\includegraphics[width=17cm]{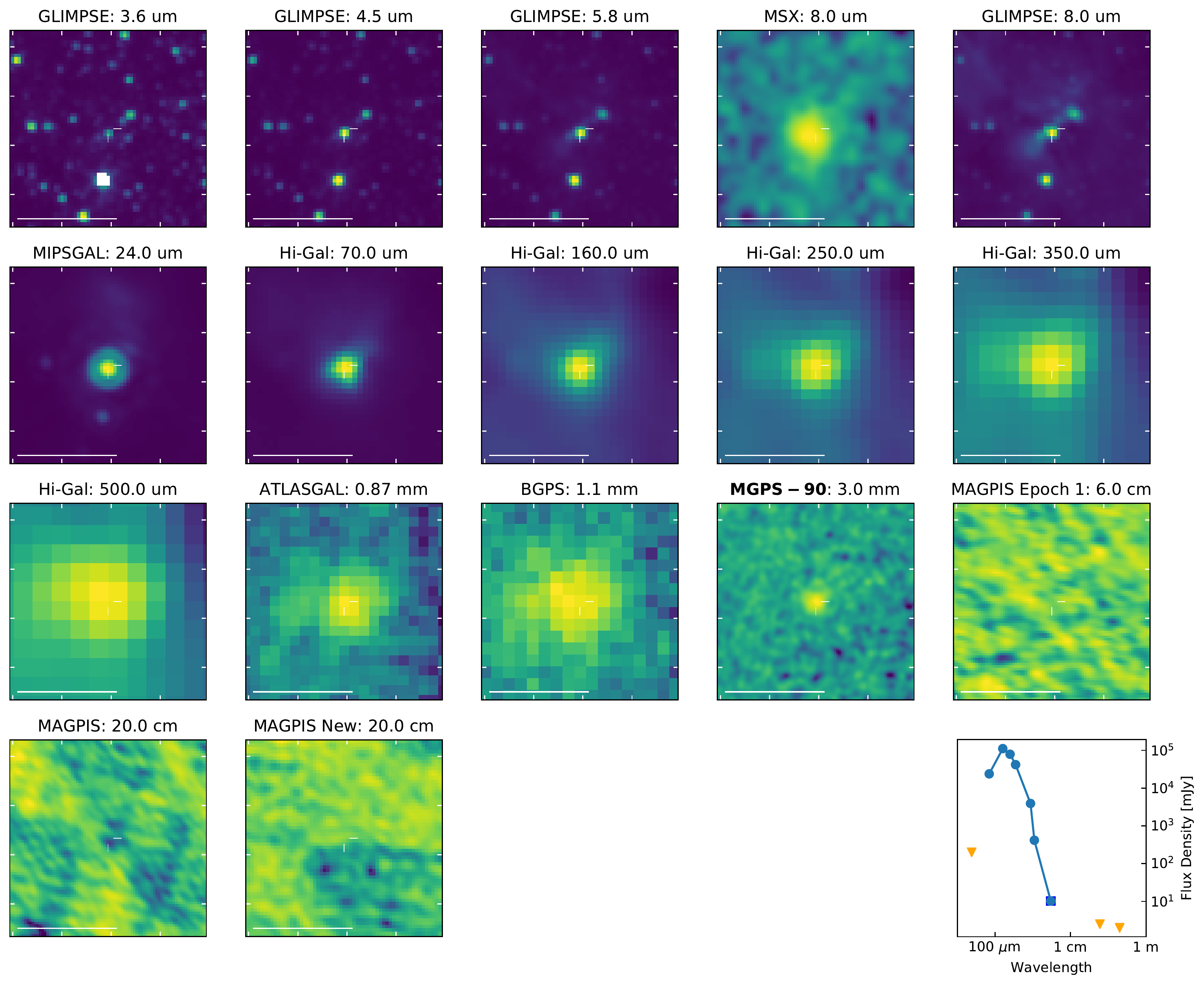}
\caption{SED of source G12.026-0.031, a candidate \hchii region and likely
high-mass YSO identified in the RMS survey \citep{Lumsden2013a}.  This compact
source failed the criteria in Section \ref{sec:hiireg}, indicating that no
clear excess of free-free emission is detected at 3 mm, and suggesting that if
an \hchii region is present, it is faint.  Such objects are of particular
interest because they constitute the most compact and likely youngest
forming high-mass stars.}
\label{fig:g12hiicand}
\end{figure*}

\begin{figure*}[htp]
\includegraphics[width=17cm]{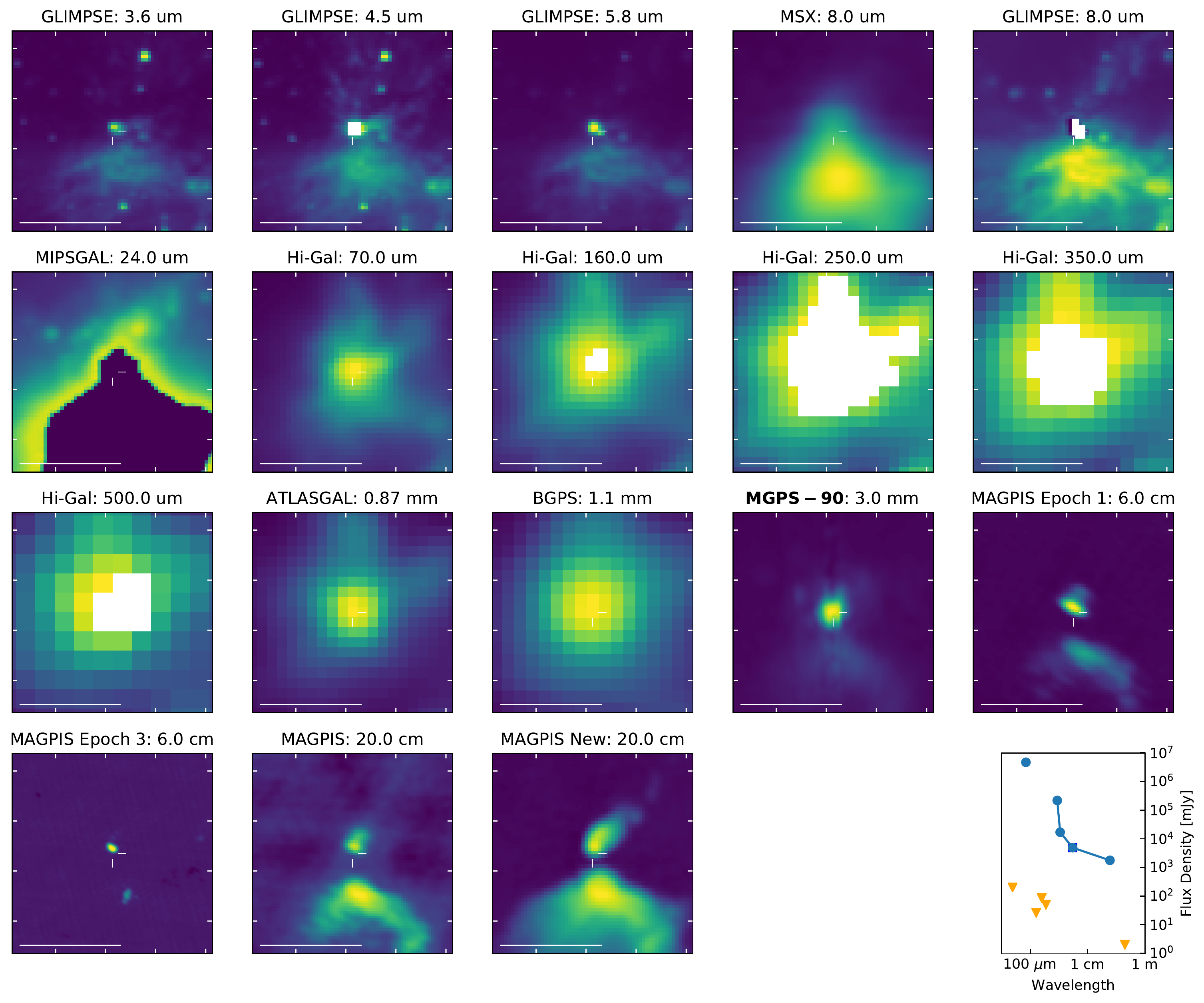}
\caption{SED of source G34.257+0.153, which contains several \hchii regions
within the MGPS90 beam \citep{Avalos2009a,Sewilo2011a}.  The source passed the
selection criteria in Section \ref{sec:hiireg}, confirming that these are
useful criteria for identifying relatively isolated \hchii regions. The
intermediate Herschel wavelengths (160, 250, 350 \um) appear as upper limits
\edit{ though the source is saturated in these Hi-Gal bands; these data points
should be ignored}.  The asymmetry in the
\MGPS image is caused by the missing scans noted in Figure
\ref{fig:g34overview}.}
\label{fig:g34hchii}
\end{figure*}

\begin{figure*}[htp]
\includegraphics[width=17cm]{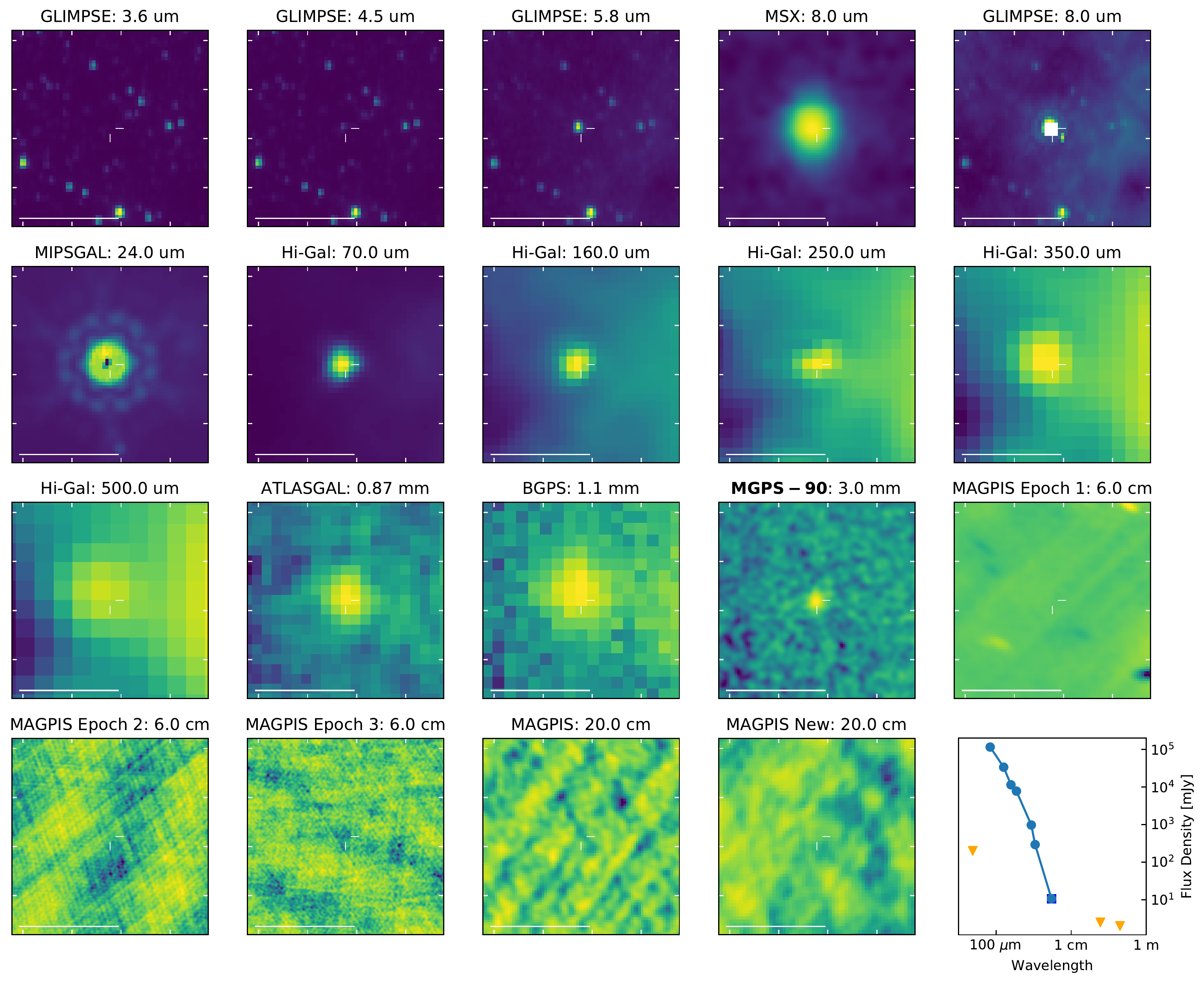}
\caption{SED of source G30.944+0.035, an OH/IR star.  The dusty SED with no
detected radio emission made this source a candidate hypercompact HII region
based on the criteria in Section \ref{sec:hiireg}, though it only barely passed
the second criterion.}
\label{fig:g31ohir}
\end{figure*}

To put the \MGPS data in context, we show a few examples of SEDs extracted
from the catalogs described in section \ref{sec:catalogmatching} along with
cutout images extracted from the same surveys.  The SEDs include the
catalog-reported flux values from each of the cross-matched surveys and the
dendrogram source flux for MUSTANG.  The selected SEDs are of a probable
planetary nebula (Fig. \ref{fig:g29pn}), which exhibits emission at all
wavelengths and was detected in extended $H-\alpha$ emission
\citep{Sabin2014a}, an OH/IR star (Fig. \ref{fig:g31ohir}) that is infrared-
and millimeter-bright but not detected at centimeter wavelengths, a high-mass
YSO that is a candidate \hchii region with no centimeter detection (Fig.
\ref{fig:g12hiicand}), and a source containing a known pair of \hchii regions
(Fig.  \ref{fig:g34hchii}).  These SEDs highlight the important role of \MGPS
data in bridging the gap between the millimeter and centimeter regimes.

\section{Diffuse Emission: Free-free separation}
\label{sec:freefree}
As stated in the introduction, the MGPS90 data have contributions from thermal
free-free, thermal dust continuum, and nonthermal synchrotron emission.  We
describe here our decomposition of the MGPS90 data into free-free and dust
emission; the non-thermal emission was not separated from the free-free
emission.

We use the ATLASGAL 870\,$\mu$m data \citep{Schuller2009a} to estimate the dust
contribution since at 870\,$\mu$m essentially all emission is from dust.  We
estimate the 90\,GHz flux density from dust by scaling the ATLASGAL data
assuming a dust emissivity index $\beta=1.5$.  Using this value of $\beta$, the
ATLASGAL 870\,$\mu$m and MGPS90 data flux densities are related via $S_{\rm
90GHz} \simeq 0.013~S_{\rm 870\mu m}$ (cf. Equation~\ref{eq:dust}).
\edit{Values of $\beta$ ranging from $1 < \beta < 2$ are often inferred from SED modeling,
so there is substantial (factor of $\sim4$) uncertainty in the extrapolated dust fluxes.
While this uncertainty limits our ability to quantitatively interpret the dust-subtracted
images, the morphology of these images is less affected.}  We
subtract the scaled ATLASGAL data from an appropriately smoothed version of the
MGPS90 map to obtain an estimated free-free map.  We perform this subtraction
on the feathered MGPS90 and Planck data (Section~\ref{sec:feather}).

Similarly, we use 20\,cm maps to estimate the dust contribution by subtracting
a scaled 20\,cm map from the MGPS90 data.
For most fields, we use 20\,cm MAGPIS data \citep{Helfand2006a}, which has an
angular resolution of $\sim\!6\arcsec$ and a point source sensitivity of
$1-2$\,mJy.  MAGPIS does not cover the Galactic center or $\ell>48^\circ$, and
so we use other data in these zones. In the Galactic center, we use the
multi-configuration 20\,cm map from \citet[][resolution
$\sim30\arcsec$]{Yusef-Zadeh2004a}, and in the W51 field we use the
multi-configuration map from \citet[][resolution $\lesssim1\arcsec$]{Mehringer1994a}.
We scale the 20\,cm to 90\,GHz assuming the 20\,cm consists exclusively of optically
thin free-free emission  following a power law $S_{\nu}
\propto \nu^{-0.12}$ \citep{Wilson2009a}.
The observed fields were selected based on their rich ongoing star formation
activity, so this approximation is reasonable, but there are several cases
where additional emission mechanisms (e.g., synchrotron) \edit{contribute to the
observed intensity.}

We show the results of the decomposition for one example field in
Figure~\ref{fig:arches_freefree}; the rest of the MGPS90 fields are in the
Appendix.  Figure~\ref{fig:arches_freefree} contains panels of the MGPS90 data,
the contribution to the MGPS90 data from thermal dust estimated from ATLASGAL
subtraction, the contribution to the MGPS90 data from free-free and synchrotron
emission, 20\,cm data, and
the contribution to the MGPS90 data from thermal dust estimated from 20\,cm
subtraction.

The example in Figure \ref{fig:arches_freefree} shows good agreement between the
two dust estimates and between the free-free estimates, and the differences
highlight some of the incorrect assumptions in the above analysis.  The excess
diffuse emission in the rightmost panel (MGPS90 - VLA) is most likely caused by
the VLA's failure to recover large angular scales.  The missing emission on the
right side of that map is caused by the excess synchrotron emission in the Sgr
A region, which is not accounted for in our simple free-free model.  Both dust
maps do well at recovering emission from the massive G0.253+0.015 cloud (the bean-shaped
feature in the upper left) and the southern dust ridge (the prominent dust feature
just below the center of the map).

The W43 region is substantially more dust-dominated than the Galactic Center
(Figure \ref{fig:w43freefree}).  The dusty features, however, are all closely
aligned with free-free features, so it is difficult to disentangle them by eye
in the MGPS90 image.  The MGPS90 - 20 cm image is negative in the 20
cm-dominated regions, likely indicating that there is substantial nonthermal
emission in these HII regions.  While there are no known supernovae in the region, the population
of OB and Wolf-Rayet stars powering the expanding HII region may also
drive strong shocks into the surrounding medium \citep[e.g.][]{Bally2010a}, leading to nonthermal emission.
The presence of such nonthermal emission indicates that electrons must be
accelerated to relativistic velocities in the HII region, which has recently
been shown to be possible in HII region expansion fronts
\citep{Padovani2019a}.

The W49B supernova remnant in the G43 field stands out as a bright nonthermal
source.  No other supernova remnants in the surveyed area are as bright at 3\,mm (see
Figure~\ref{fig:g43overview}). \citet{Sun2011a} found that the spectral energy
distribution of W49B is well-fit by a single power law from 200\,MHz to 30\,GHz
with index $\alpha=-0.46\pm0.01$.  They found that the 5\,GHz integrated flux density is
$19.10\pm0.98$\,Jy, and the 90\,GHz integrated flux density should be 5.1\,Jy.
Integrating over W9B, we find a flux density of 5.2\,Jy, indicating that most of
the associated emission is nonthermal. 

The decomposed images are shown in Figures
\ref{fig:arches_freefree}-\ref{fig:w51mainfreefree}.

\begin{table*}[htp]
\centering
\caption{Free-free and Dust Emission Fractions}
\begin{tabular}{llll}
    \label{tab:freefree}
Field Name   & Center (Galactic) & Field Size & Dust Fraction ($\beta=1.5$) \\
             & degrees           & arcminutes &                             \\
\hline
\hline
Arches     & 0.140 -0.054                        & 12.60 &        0.08 \\
Sgr\,B2    & 0.657 -0.041                        & 12.80 &        0.13 \\
W33        & 12.805 -0.206                       &  3.72 &        0.11 \\
G29        & 29.927 -0.041                       &  5.87 &        0.11 \\
W43        & 30.757 -0.045                       &  7.24 &        0.10 \\
G34        & 34.257 0.148                        &  2.80 &        0.21 \\
W49a       & 43.171 -0.006                       &  4.38 &        0.13 \\
W49b       & 43.268 -0.186                       &  3.53 &        0.04 \\
W51a       & 49.461 -0.368                       &  8.75 &        0.14 \\
W51b       & 49.080 -0.338                       & 11.30 &        0.07 \\
\hline
\hline
\end{tabular}
\end{table*}

We directly quantify the dust contribution to the 3 mm intensity in the
targeted brightest fields.  Each field includes one or more prominent extended
structures that were the focus for these pilot observations.  For each of these
structures, we extracted an area that encompasses the bulk of the 3 mm emission
and measured the fraction of that emission that is explained by optically-thin
dust, which is the sum of the positive values from the scaled ATLASGAL data
explained above divided by the sum of the 3 mm emission.  The results are reported
in Table \ref{tab:freefree}.  Because we have assumed $\beta=1.5$, and typical
dust $\beta$ values for the ISM are $\sim1.5-2$ \citep[e.g.,][]{Ossenkopf1994a},
these can be treated as upper limits on the dust contribution.

In the regions of interest, the dust contribution at 3 mm is limited to
$\lesssim20\%$ on the several arcminute scales probed.  Regions with substantial
synchrotron contributions from supernova remnants (W49b, W51b) or other mechanisms
(the Arches) have an even lower contribution from dust, $<10\%$.  In short, the 
integrated diffuse emission detected in MGPS90 is dominated by emission from hot gas rather
than from cold molecular gas.  We are, however, unable to determine whether the \emph{area}
of the survey is dominated by hot or cold gas, as the large angular scale filtering of the
interferometric data sets prevents such an assessment; it remains possible that the area
(and volume) of the surveyed regions is dominated by cold dust emission, while the received
flux is clearly dominated by hot gas.

\begin{figure*}[htp]
    \includegraphics[width=17cm]{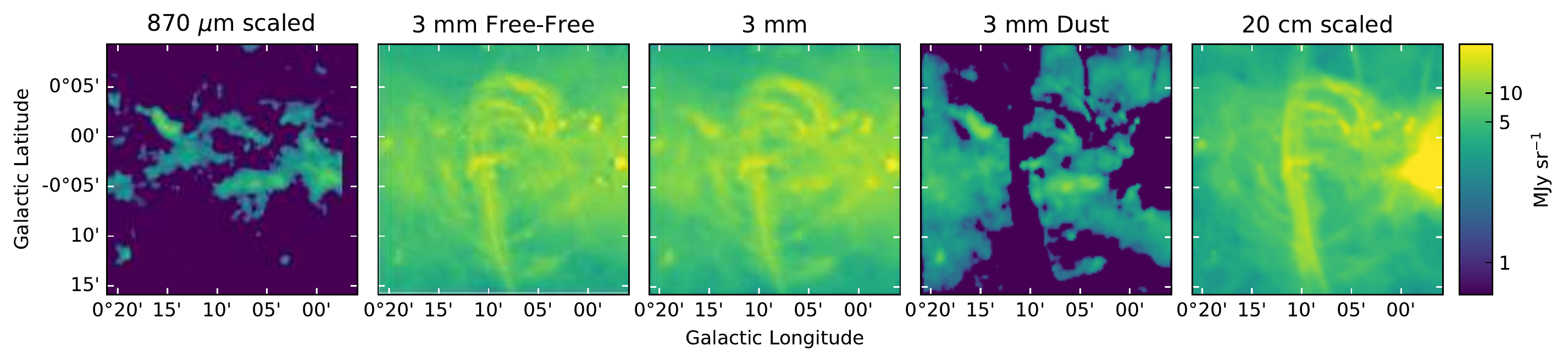}
    \caption{Decomposition of the MGPS90 data in the G01 field centered on the
    Arches region. 
    All images are displayed on the same intensity scale.  In the G01 field,
    the 20\,cm data have 30\arcsec resolution, so the MGPS90 data have been
    smoothed to match
    the resolution of the other images.
    \edit{From left to right,}
    (a; 870 \um scaled) ATLASGAL 870 \um scaled to 3 mm,
    (b; 3 mm free-free) smoothed MGPS90 - scaled ATLASGAL,
    (c; 3 mm) MGPS90,
    (d; 3 mm Dust) smoothed MGPS90 - scaled VLA 20 cm continuum,
    (e; 20 cm scaled) VLA 20 cm continuum scaled to 3 mm.
    \edit{The images are displayed with a logarithmic stretch.}
}
\label{fig:arches_freefree}
\end{figure*}

\begin{figure*}[htp]
    \includegraphics[width=17cm]{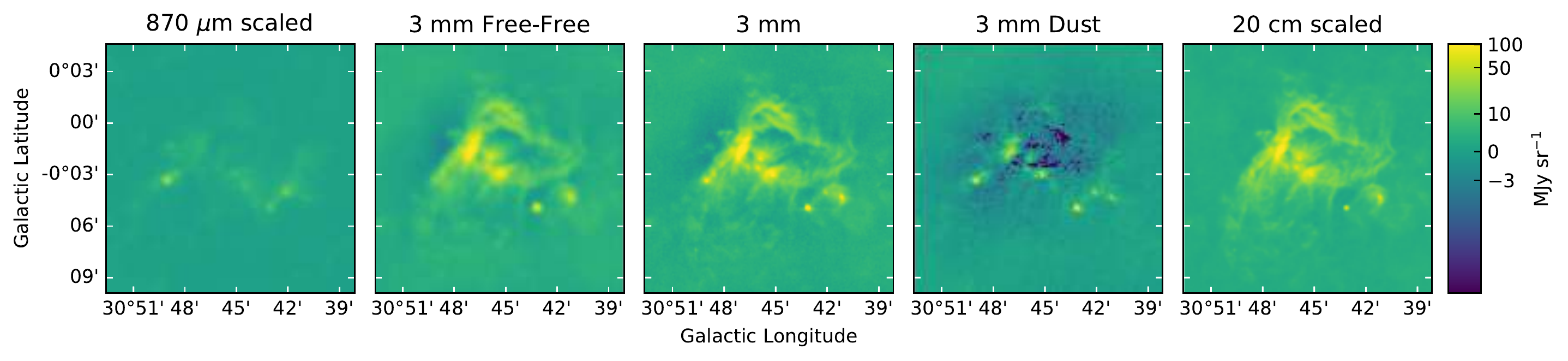}
    \caption{Decomposition of the MGPS90 data in the G31 field centered on W43.
    See Figure \ref{fig:arches_freefree} for a description of the panels.
    In G31, the 20 cm data have $\sim5$ \arcsec resolution, so they are
    smoothed to match the MGPS90 data to create panels (d) and (e), while the
    MGPS90 data are smoothed to match the ATLASGAL data to create panel (b).
}
\label{fig:w43freefree}
\end{figure*}

\section{Conclusions}
We have presented the pilot data for the MUSTANG 90 GHz Galactic Plane
Survey, MGPS90.  When complete, this survey
will cover most of the northern Galactic plane within $|b|<0.5^\circ$.
These initial data cover several high-mass star cluster
forming regions.  All imaged regions are dominated by free-free and synchrotron
emission at 3 mm.  

We cataloged emission in the images, identifying \noksources sources using the
\texttt{dendrogram} algorithm, of which \ncompacthiicand are verified \hchii
regions, and another \mmdetectionscmnondetectionscompact are plausible
candidates.

\acknowledgements
We thank the anonymous referee for a detailed and constructive report.
\MUSTANG is funded by the NSF award number 1615604 and by the Mt.\ Cuba
Astronomical Foundation. The National Radio Astronomy Observatory is a facility
of the National Science Foundation operated under cooperative agreement by
Associated Universities, Inc. 
RGM acknowledges support from UNAM-PAPIIT project IN104319.

\appendix

\section{Additional free-free / dust decomposition maps}
In this appendix, we show cutout images focused on a selection of bright
extended emission regions and the associated free-free decomposition described
in Section \ref{sec:freefree}.

\begin{figure*}[htp]
    \includegraphics[width=17cm]{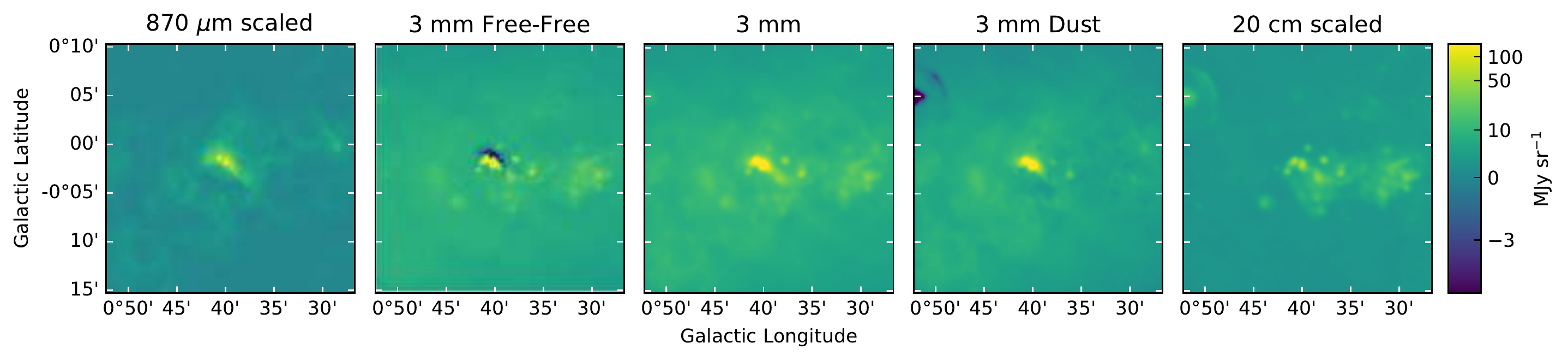}
    \caption{Decomposition of the MGPS90 data in the G01 field centered on Sgr B2.
    See Figure \ref{fig:arches_freefree}.
    The differences in the ATLASGAL- and 20 cm-based dust decomposition highlight the
    different angular scales recovered by those data sets.
}
\label{fig:sgrb2freefree}
\end{figure*}

\begin{figure*}[htp]
    \includegraphics[width=17cm]{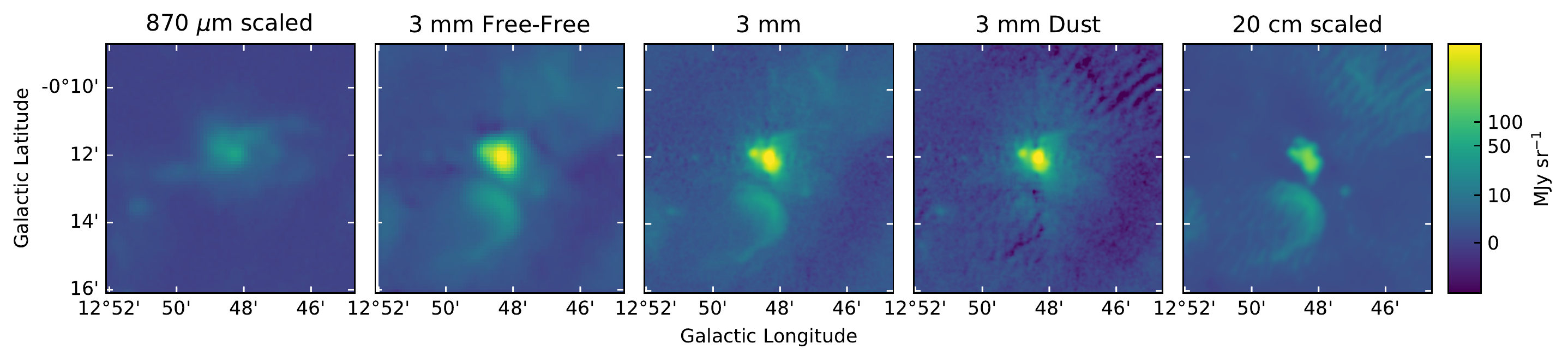}
    \caption{Decomposition of the MGPS90 data in the G12 field centered on W33.
    See Figure \ref{fig:arches_freefree} for a description of the panels.
    The diffuse free-free emission is well-removed by subtracting the 20 cm data,
    but the compact point source appears much brighter in the 3 mm-derived map;
    this difference is likely because free-free emission is present but optically
    thick at 20 cm, resulting in an underestimate of the free-free contribution 
    at 3 mm.
}
\label{fig:w33freefree}
\end{figure*}

\begin{figure*}[htp]
    \includegraphics[width=17cm]{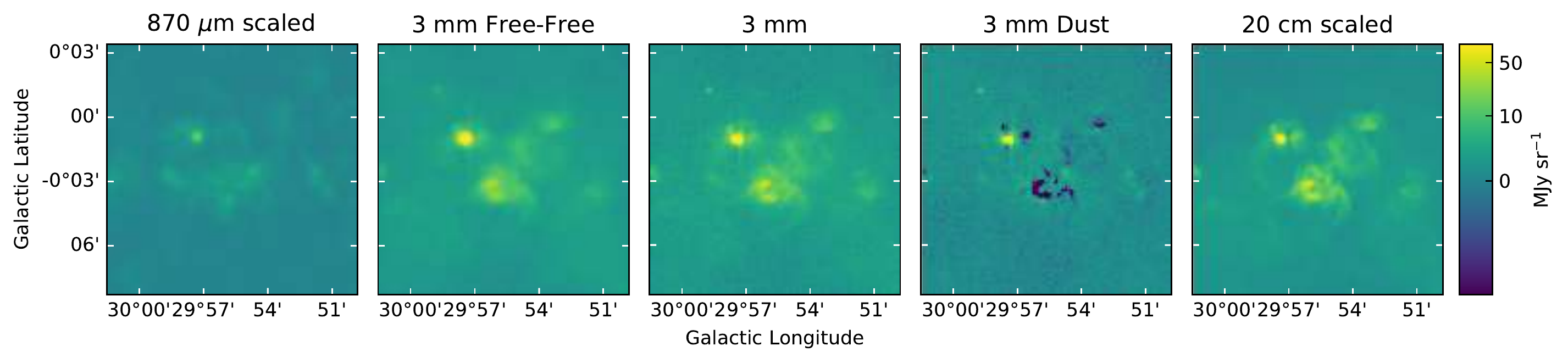}
    \caption{Decomposition of the MGPS90 data in the G29 field.
    See Figure \ref{fig:arches_freefree} for a description of the panels.
    Some of the compact structures exhibit strong excesses at 20 cm.
}
\label{fig:g29freefree}
\end{figure*}

\begin{figure*}[htp]
    \includegraphics[width=17cm]{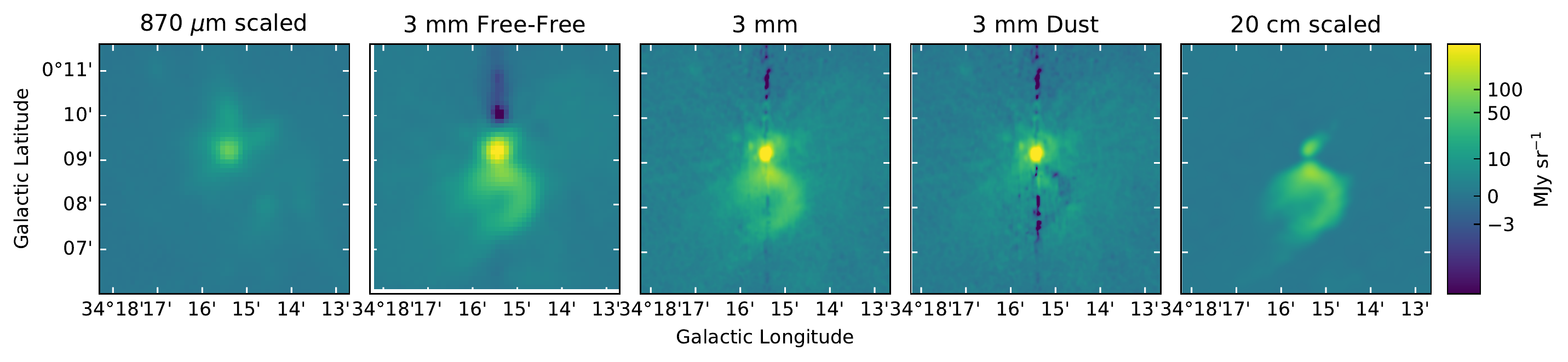}
    \caption{Decomposition of the MGPS90 data in the G34 field centered on G34.26+0.15.
    See Figure \ref{fig:arches_freefree} for a description of the panels.
    The vertical streak is an artifact as mentioned in Figure \ref{fig:g34overview}.
}
\label{fig:g34freefree}
\end{figure*}

\begin{figure*}[htp]
    \includegraphics[width=17cm]{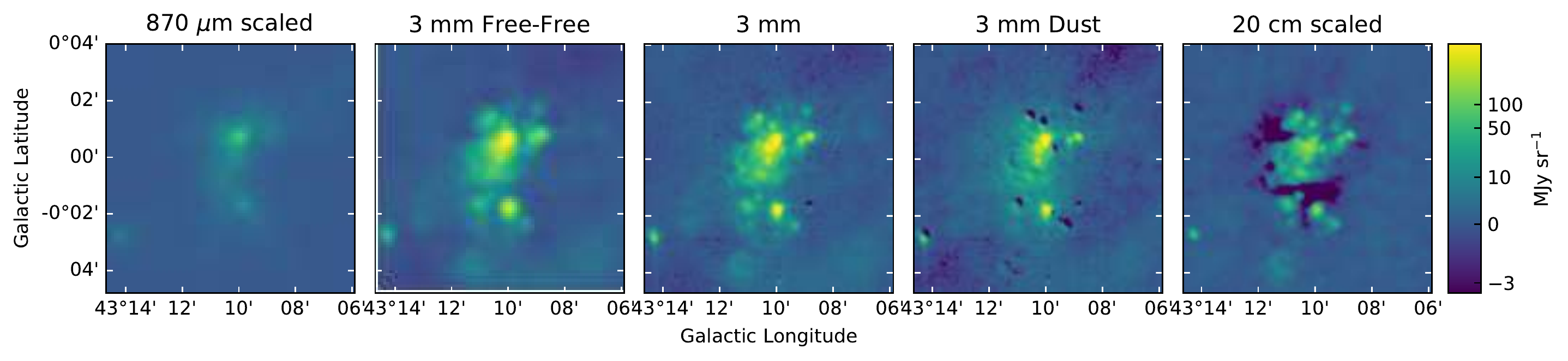}
    \caption{Decomposition of the MGPS90 data in the G43 field centered on W49A.
    See Figure \ref{fig:arches_freefree} for a description of the panels.
    As in Figure \ref{fig:g29freefree}, several compact structures appear to have
    excess 20 cm emission.  However, other structures exhibit free-free
    emission that is optically thick at 20 cm and is therefore under-subtracted
    at 3 mm in panel (d); see also Figure \ref{fig:w51mainfreefree}.
}
\label{fig:w49afreefree}
\end{figure*}

\begin{figure*}[htp]
    \includegraphics[width=17cm]{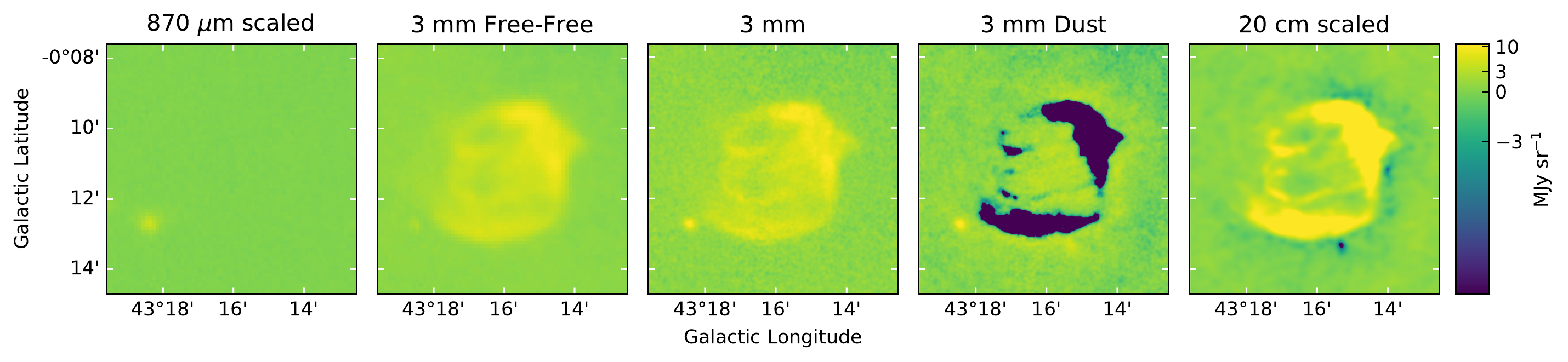}
    \caption{Decomposition of the MGPS90 data in the G43 field centered on W49B.
    See Figure \ref{fig:arches_freefree} for a description of the panels.
    W49B is a supernova remnant completely dominated by synchrotron emission.
    Panel (b) therefore shows synchrotron, not free-free, emission.
}
\label{fig:w49bfreefree}
\end{figure*}

\begin{figure*}[htp]
    \includegraphics[width=17cm]{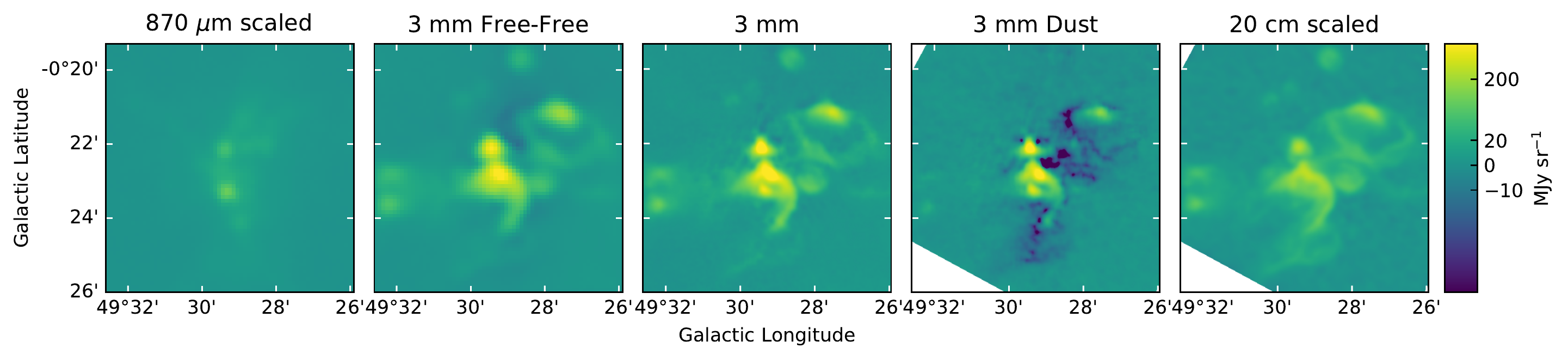}
    \caption{Decomposition of the MGPS90 data in the G49 field centered on W51 Main.
    See Figure \ref{fig:arches_freefree} for a description of the panels.
    There is a mix of under- and over-subtracted emission in the dust map in panel (d);
    the arc shape in the center is purely free-free emission \citep{Ginsburg2016a,Ginsburg2017a}, but
    it is optically thick at 20 cm.
}
\label{fig:w51mainfreefree}
\end{figure*}

\section{Catalog}
\label{appendix:Catalog}
Table \ref{tab:photometry} shows an excerpt from the catalog including
the brightest 20 sources.  \edit{The full catalog will be published
electronically with the paper.}
We include the dendrogram measurements of the integrated flux density and
the Galactic $\ell$ and $b$ centroids, integrated flux densities in 10\arcsec
and 15\arcsec apertures, the median background in a 15-20\arcsec aperture,
and the parameters of the best-fit two-dimensional Gaussian profile.
The sample table is sorted by the Gaussian peak amplitude ($A_G$) in descending
order.

\begin{table}[htp]
\caption{\MUSTANG Source IDs and photometry}
\resizebox{\textwidth}{!}{
\begin{tabular}{lllllllllllll}
\label{tab:photometry}
ID & Dendrogram $S_{\nu}$ & $\ell$ & $b$ & $S_{\nu,10''}$ & $S_{\nu,15''}$ & $S_{bg;15-20''}$ & $A_G$ & $\ell_G$ & $b_G$ & FWHM$_{maj,G}$ & FWHM$_{min,G}$ & PA$_G$ \\
 & $\mathrm{Jy}$ & $\mathrm{{}^{\circ}}$ & $\mathrm{{}^{\circ}}$ & $\mathrm{Jy}$ & $\mathrm{Jy}$ & $\mathrm{Jy\,beam^{-1}}$ & $\mathrm{Jy\,beam^{-1}}$ & $\mathrm{{}^{\circ}}$ & $\mathrm{{}^{\circ}}$ & $\mathrm{{}^{\prime\prime}}$ & $\mathrm{{}^{\prime\prime}}$ & $\mathrm{{}^{\circ}}$ \\
\hline
35.00 & 17.01 & 12.805 & -0.201 & 3.657 & 6.887 & 2.292 & 5.80 & 12.806 & -0.201 & 17.034 & 14.966 & 90 \\
56.00 & 8.30 & 43.167 & 0.010 & 3.099 & 5.49 & 1.564 & 5.22 & 43.167 & 0.010 & 15.685 & 11.546 & 76.062 \\
166.00 & 12.58 & 0.668 & -0.035 & 2.699 & 5.067 & 1.846 & 3.94 & 0.668 & -0.036 & 16.34 & 15.036 & 343.275 \\
14.00 & 4.88 & 34.257 & 0.153 & 2.333 & 3.728 & 0.592 & 4.56 & 34.257 & 0.153 & 11.208 & 9.439 & 67.04 \\
66.00 & 6.43 & 49.492 & -0.368 & 2.071 & 3.644 & 1.059 & 3.10 & 49.492 & -0.368 & 16.674 & 13.335 & 132.052 \\
49.00 & 11.59 & 49.489 & -0.380 & 1.698 & 3.408 & 1.481 & 2.06 & 49.488 & -0.380 & 21.262 & 14.499 & 141.889 \\
42.00 & 4.67 & 43.166 & -0.030 & 1.394 & 2.501 & 0.773 & 2.33 & 43.166 & -0.030 & 13.947 & 13.589 & 360 \\
142.00 & 4.16 & 359.946 & -0.046 & 0.853 & 1.604 & 0.624 & 1.42 & 359.946 & -0.046 & 14.779 & 14.363 & 360 \\
116.00 & 2.70 & 29.957 & -0.017 & 0.766 & 1.451 & 0.409 & 1.37 & 29.956 & -0.017 & 14.89 & 12.647 & 65.432 \\
36.00 & 0.95 & 12.813 & -0.199 & 0.692 & 1.19 & 0.345 & 1.15 & 12.812 & -0.199 & 27.0 & 21.545 & 360 \\
41.00 & 2.64 & 29.957 & -0.018 & 0.653 & 1.204 & 0.402 & 1.06 & 29.957 & -0.018 & 15.584 & 14.434 & 0 \\
45.00 & 1.40 & 49.491 & -0.386 & 0.602 & 1.175 & 0.428 & 0.74 & 49.491 & -0.386 & 27.0 & 19.286 & 256.306 \\
51.00 & 0.69 & 43.172 & -0.001 & 0.448 & 0.82 & 0.251 & 0.70 & 43.172 & -0.000 & 16.815 & 15.165 & 360 \\
179.00 & 1.32 & 31.412 & 0.308 & 0.374 & 0.696 & 0.201 & 0.61 & 31.412 & 0.307 & 15.161 & 13.027 & 121.242 \\
164.00 & 0.83 & 30.866 & 0.114 & 0.372 & 0.593 & 0.101 & 0.69 & 30.866 & 0.114 & 11.418 & 10.267 & 129.634 \\
62.00 & 0.93 & 30.720 & -0.083 & 0.362 & 0.617 & 0.136 & 0.68 & 30.720 & -0.083 & 12.342 & 11.126 & 130.278 \\
55.00 & 1.85 & 43.149 & 0.012 & 0.326 & 0.623 & 0.241 & 0.73 & 43.148 & 0.013 & 19.1 & 13.303 & 0 \\
156.00 & 0.57 & 0.658 & -0.042 & 0.261 & 0.447 & 0.109 & 0.48 & 0.659 & -0.041 & 27.0 & 13.659 & 135.233 \\
133.00 & 0.90 & 30.534 & 0.021 & 0.248 & 0.446 & 0.132 & 0.41 & 30.534 & 0.021 & 15.429 & 12.704 & 150.774 \\
\hline
\end{tabular}
}\par
The subscripts X${_G}$ are for the parameters derived from Gaussian fits.  The values displayed are rounded such that the error is in the last digit; error estimates can be found in the digital version of the table.Note that position angles in the set (0, 90, 180, 270, 360) are caused by bad fits.  These fits are kept in the catalog because they passed other criteria and are high signal-to-noise, but they are likely of sources in crowded regions so the corresponding fit parameters should be treated with caution.
\end{table}

\end{document}